\documentclass[12pt]{article}

\RequirePackage{cite}
\RequirePackage{times}
\RequirePackage{fullpage}
\RequirePackage{ifthen}

\usepackage{enumerate}

\usepackage[usenames,dvipsnames]{color}
\usepackage{xcolor}

\makeatletter
\def\@cite#1#2{$^{\mbox{\scriptsize #1\if@tempswa , #2\fi}}$}
\makeatother

\newcommand{\spacing}[1]{\renewcommand{\baselinestretch}{#1}\large\normalsize}
\spacing{2}

\def\@maketitle{%
  \newpage\spacing{1}\setlength{\parskip}{12pt}%
    {\Large\bfseries\noindent\sloppy \textsf{\@title} \par}%
    {\noindent\sloppy \@author}%
}

\newenvironment{affiliations}{%
    \setcounter{enumi}{1}%
    \setlength{\parindent}{0in}%
    \slshape\sloppy%
    \begin{list}{\upshape$^{\arabic{enumi}}$}{%
        \usecounter{enumi}%
        \setlength{\leftmargin}{0in}%
        \setlength{\topsep}{0in}%
        \setlength{\labelsep}{0in}%
        \setlength{\labelwidth}{0in}%
        \setlength{\listparindent}{0in}%
        \setlength{\itemsep}{0ex}%
        \setlength{\parsep}{0in}%
        }
    }{\end{list}\par\vspace{12pt}}

\renewenvironment{abstract}{%
    \setlength{\parindent}{0in}%
    \setlength{\parskip}{0in}%
    \bfseries%
    }{\par\vspace{-4pt}}

\newenvironment{addendum}{%
    \setlength{\parindent}{0in}%
    \small%
    \begin{list}{Acknowledgements}{%
        \setlength{\leftmargin}{0in}%
        \setlength{\listparindent}{0in}%
        \setlength{\labelsep}{0em}%
        \setlength{\labelwidth}{0in}%
        \setlength{\itemsep}{12pt}%
        }
    }
    {\end{list}\normalsize}

\newcommand*{\addendumlabel}[1]{\textbf{#1}\hspace{1em}}

\usepackage{times}
\usepackage{subfigure}

\usepackage{epstopdf}
\usepackage{graphicx}
\usepackage{epsfig}

\newcommand{\be}{\begin{equation}}
\newcommand{\ee}{\end{equation}}

\def \aap{{\it Astron.~Astrophys}}
\def \apjl{{\it Astrophys.~J.~Lett.}}
\def \apj{{\it Astrophys.~J.}}
\def \apjs{{\it Astrophys.~J. Suppl.~S.}}
\def \araa{{\it Annu.~Rev.~Astron.~Astr.}}

\def \mnras{{\it Mon.~Not.~Roy.~Astron.~Soc.}}
\def \prl{{\it Phys.~Rev.~Lett.}}

\def \nat{{\it Nature}}

\def \aj{{\it Astron.~J.}}

\def \pasp{{\it Publ.~Astron.~Soc.~Pac.}}%irs just put this in as a fix, not sure if this is right
\def\msun{{\,M_\odot}}

\def\lesssim{\lower.5ex\hbox{$\; \buildrel < \over \sim \;$}}
\def\gtrsim{\lower.5ex\hbox{$\; \buildrel > \over \sim \;$}}
\newcommand{\lsim}{\,\rlap{\raise 0.35ex\hbox{$<$}}{\lower 0.7ex\hbox{$\sim$}}\,}
\newcommand{\gsim}{\,\rlap{\raise 0.35ex\hbox{$>$}}{\lower 0.7ex\hbox{$\sim$}}\,}

\newcommand{\bea}{\begin{eqnarray}}
\newcommand{\eea}{\end{eqnarray}}

\begin{document}

\noindent
 {\large {\bf {\fontfamily{phv}\selectfont Diffuse Galactic Antimatter from Faint Thermonuclear Supernovae  in Old Stellar Populations
}}}

\noindent
Roland~M.~Crocker$^{1}$, %\thanks{E-mail: rcrocker@fastmail.fm}\thanks{Australian Research Council Future Fellow}, 
Ashley~J.~Ruiter$^{1,2}$,
Ivo~R.~Seitenzahl$^{1,2,3}$,
Fiona~H.~Panther$^{1,2}$,
Stuart Sim$^{4}$,
Holger Baumgardt$^{5}$,
Anais M{\"o}ller$^{1,2}$,
David~M.~Nataf$^{1,6,7}$,
Lilia Ferrario$^{8}$,
%Jarrod Hurley${^5}$,
J.J. Eldridge$^{9}$,
Martin White$^{10}$,
Brad~E.~Tucker$^{1,2}$,
and Felix Aharonian$^{11,12}$

\begin{affiliations}
 \item Research School of Astronomy and Astrophysics, Australian National University, Canberra, Australia
  \item ARC Centre of Excellence for All-sky Astrophysics (CAASTRO), Canberra, Australia
 \item School of Physical, Environmental and Mathematical Sciences, UNSW Canberra, Australian Defence Force Academy, Canberra, Australia
 \item School of Mathematics and Physics, Queen's University, Belfast, U.K.
\item School of Mathematics and Physics, University of Queensland, Brisbane, Australia
\item Department of Physics and Astronomy,
The Johns Hopkins University, Baltimore, U.S.A.
 \item Center for Astrophysical Sciences and Department of Physics and Astronomy, The Johns Hopkins University, Baltimore, MD 21218.
 \item Mathematical Sciences Institute, Australian National University, Canberra, Australia
 %\item Centre for Astrophysics and Supercomputing, Swinburne University of Technology, Melbourne, Australia
 \item Department of Physics, University of Auckland, Auckland, New Zealand
\item Department of Physics, University of Adelaide, Adelaide, Australia
\item Dublin Institute for Advanced Studies, Dublin 2, Ireland
\item Max-Planck-Institut f{\" u}r Kernphsik, Heidelberg, Germany
\end{affiliations}

%%%%%%%%%%%%%%%%%%%%%%%%%%%%%%%%%%%%%%%%%%%%%%%%%%%%%%%%%%%%%%%
%%%%%%%%%%%%%%%%%%%%%%%%%%%%%%%%%%%%%%%%%%%%%%%%%%%%%%%%%%%%%%%

\begin{abstract}
Our Galaxy hosts the annihilation of a few $\times 10^{43}$ low-energy positrons every second.
Radioactive isotopes capable of supplying such positrons are synthesised in stars, stellar remnants, and supernovae.
For decades, however,
there has been no positive identification of a main stellar positron source leading to
suggestions that many  positrons originate 
from exotic sources like the Galaxy's central super-massive black hole
or dark matter annihilation. %, but such sources would not explain the recently-detected positron signal from the extended Galactic disk.
Here we show that a single type of transient source, 
deriving from stellar populations of age 3-6 Gyr and yielding  $\sim 0.03 \msun$ of the positron emitter $^{44}$Ti,
can simultaneously explain the strength  and morphology of the Galactic positron annihilation signal
and the solar system abundance of the $^{44}$Ti decay product $^{44}$Ca.
This transient is likely the merger of two low-mass white dwarfs,
observed in external galaxies as the sub-luminous, thermonuclear supernova known as SN1991bg-like.
\end{abstract}

First detected more than forty years ago\cite{Johnson1972},
the Galactic positron ($e^+$) annihilation signal poses two  central, unresolved challenges: 
i) to understand the absolute positron production rate in the Galaxy and 
ii) to explain the gross morphology of the positron annihilation distribution across the Galaxy.
In particular, historical measurements\cite{Prantzos2011} have suggested a Galactic bulge\cite{SI} positron 
annihilation rate $\sim$ 1.4 times larger than the Galactic disc annihilation rate, despite the bulge's hosting  $\lsim 1/2$ the stellar mass\cite{Bland-Hawthorn2016} and much less ($\lsim 1/10$) 
recent star formation than the disc.

This apparent strong bulge emission spurred the development of models 
wherein  positrons are injected by  `exotic' sources like dark matter annihilation or the central super-massive black hole.
However, severe gamma-ray continuum constraints at energies above 511 keV[\cite{ Aharonian1981,Beacom2006}] 
imply that most Galactic positrons are injected into the interstellar medium (ISM) with kinetic energies $\lesssim 3$ MeV.
This has ruled out many  explanations for positron origin involving dark matter and others invoking diffuse, hadronic cosmic rays or compact positron sources like pulsars\cite{Prantzos2011}.
The energy constraint also means
that  the positrons' diffusive transport distance is $\lsim$kpc[\cite{ Alexis2014}] %Martin2012}]
within their $10^5-10^6$ year[\cite{Churazov2011}] 
ISM lifetimes. %(so a SMBH origin for the $\sim 3$ kpc radius bulge positron annihilation radiation would require non-diffusive positron transport out of the nucleus\cite{Crocker2011}).
The positrons are therefore expected to annihilate relatively close to their injection sites, meaning that the Galaxy presents a thick target
to the positrons on these spatial scales.
Given we also expect the $e^+$ annihilation in the Galaxy to be in quasi steady state 
(because the time between $e^+$ injection events is much less than the $e^+$ ISM lifetimes), Galactic $e^+$ production is in saturation.
This implies that i) the sky distribution of annihilation radiation broadly reflects the distribution of positron sources; 
and ii) the current 
positron injection rate into the ISM is equal to the annihilation rate inferred from the annihilation radiation flux.

Recently the empirical situation regarding positron annihilation has undergone two important updates following a novel analysis\cite{Siegert2015}  of a larger data set generated by the SPI spectrometer\cite{Roques2003} on ESA's INTEGRAL satellite: 
i) the measured disc positron annihilation rate has been subject to a significant upwards revision following the detection of considerably more low surface brightness emission 
(implying a revised total Galactic positron annihilation rate $5.0_{-1.5}^{+1.0} \times 10^{43}$ s$^{-1}$); 
and
ii) the existence of a distinct, point-like, Galactic centre source has been demonstrated with $5\sigma$ statistical significance (also see \cite{Skinner2014}).

These new findings have significant consequences for our understanding of Galactic positron production.
Given the revision of the measured disc annihilation rate, the bulge to disc positron luminosity ratio is revised downwards to $B/D = 0.42 \pm 0.09$; this is equal, within errors, to the ratio of bulge to disc stellar mass\cite{Bland-Hawthorn2016,SI}: $M_\mathrm{bulge}/M_\mathrm{disc} = (1.6\pm0.2)\times 10^{10} \msun/(3.7\pm0.5)\times 10^{10} \msun = 0.4 \pm 0.1$.
The effective angular resolution of the SPI observations of $2.7^\circ$[\cite{Siegert2015}] means the point-like Galactic centre positron source 
encompasses the entire `nuclear bulge' \cite{Launhardt2002}  stellar population surrounding the supermassive black hole.
The ratio of the positron luminosities of the nuclear bulge  to the bulge is $N/B = (8.3 \pm 2.1) \times 10^{-2}$[\cite{Siegert2015}], 
again statistically equal to the ratio of the 
stellar masses of these structures: $M_\mathrm{NB}/M_\mathrm{bulge} = (1.4\pm0.6)\times 10^{9} \msun/(1.6\pm0.2)\times 10^{10} \msun = 0.09 \pm 0.04$.
Together, these facts imply that a single type of positron source connected to old stars could explain the global distribution of positron injection in the Galaxy
(in contrast, putative `exotic' sources associated with the inner Galaxy, like dark matter or the central super-massive black hole, cannot explain the disk positron signal).

Qualitatively, given the strong bulge signal, it has long been appreciated that a positron source connected to old stellar populations 
seems to be preferred.
For instance, ordinary Type Ia (thermonuclear) supernovae (SNe~Ia) have been a theoretically-favoured, putative source of Galactic positrons \cite{Higdon2009,Prantzos2011}.
Alternatively, following older suggestions\cite{Guessoum2006}
it was recently noted\cite{Siegert2016} that flaring microquasars, also plausibly connected to old stars in some instances, seem capable of sustaining the Galactic $e^+$ annihilation rate.

\begin{figure}%%[hb!]
\centering
\includegraphics[width = 1 \textwidth]{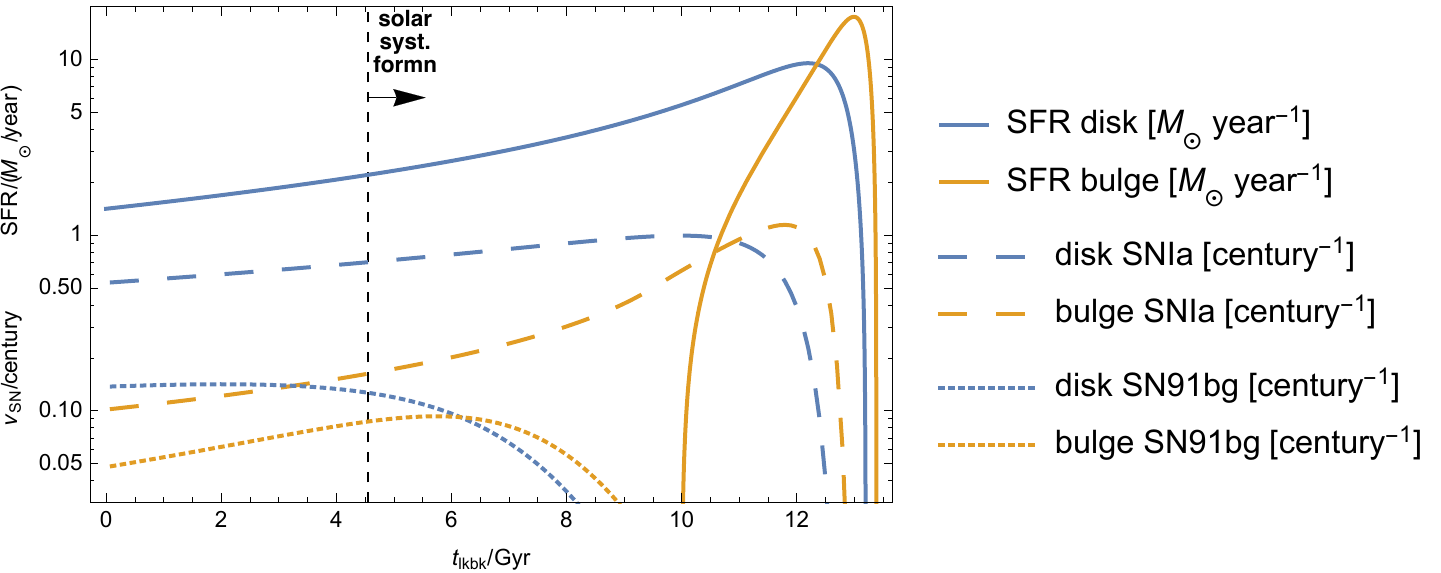}
\caption{{\bf The star formation rates (SFRs) of the disc and bulge adopted in this work and the resulting SN~Ia and SN91bg rates} 
plotted as function of lookback time (measured backward from the present).
The SFRs (solid curves) are plotted in $\msun$ year$^{-1}$; the SN rates (dotted and dashed) are shown in century$^{-1}$.
The SFRs are constrained to integrate (once stellar mass loss is accounted for) to the existing stellar masses of the various regions and to match their stellar age distributions\cite{SI}.
For stellar masses we adopt\cite{Bland-Hawthorn2016}: a current total stellar mass of the Milky Way of $(5.2 \pm 0.5) \times 10^{10} \msun$,
total mass in the bulge volume of $(1.6 \pm 0.2) \times 10^{10} \msun$ and a stellar mass
in the disc outside the bulge volume of $(3.7 \pm 0.5) \times 10^{10} \msun$.
The star formation rate in the bulge and disc both peaked 12-13 Gyr ago but, setting aside the nuclear bulge, the bulge has experienced negligible star formation since 
about 10 Gyr ago\cite{Nataf2015}, whereas the disc has continued to host star formation at a rate $\sim 1-2 \msun$ year$^{-1}$ up until the present.
Note that the disc SN91bg rate is the only illustrated rate currently increasing with cosmic time
(the nuclear bulge SN91bg rate, not shown here, is also currently increasing).
The SN rates are calculated using the formalism of delay time distributions (see caption to Fig.~\ref{fig_plotDTDforPaper}) with $t_p = 0.3$ Gyr and $s = -1.0$ for SNe~Ia and $t_p = 5.4$ Gyr and $s = -1.6$ for SN91bg.
}
\label{fig_plotSFRMW}
\end{figure}

\begin{figure}%%[hb!]
\centering
\includegraphics[width = 1 \textwidth]{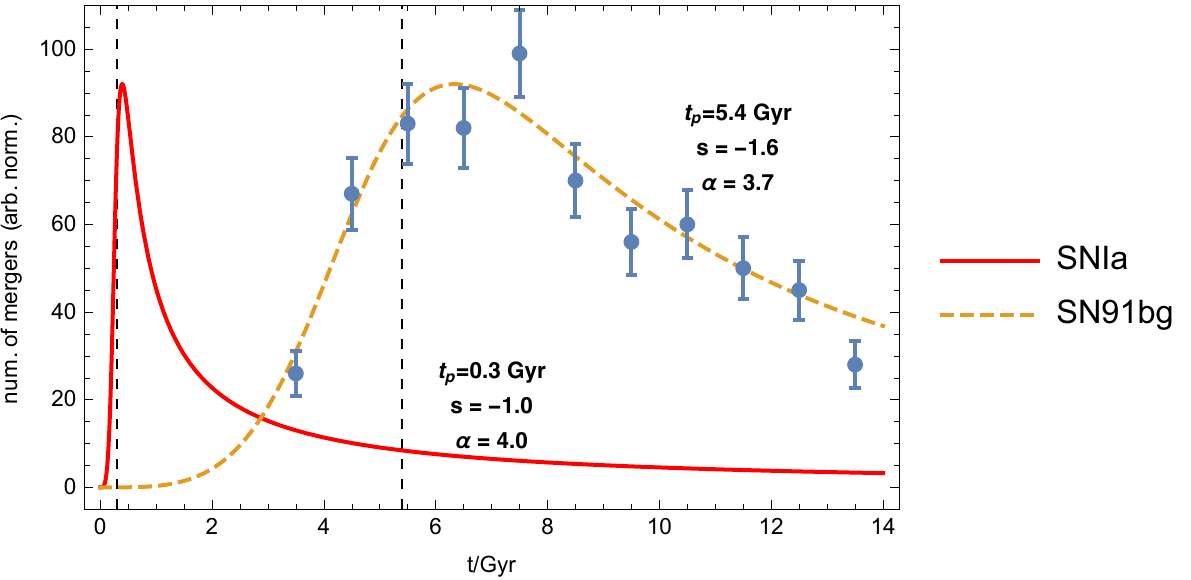}
\caption{{\bf Binned CO-WD He WD mergers within our BPS model} (data points with error bars given by $\pm \sqrt{N}$)
together with delay time distribution (DTD) curves for SNe Ia  and CO-WD He WD mergers (identified with SNe 91bg) plotted as 
a function of  time (increasing from the star formation event towards the present); the DTD curves are arbitrarily scaled in the vertical direction.
The DTD returns the rate of events of type $X$ 
at epoch $t$ (measured since the Big Bang) when convolved with the previous star formation history (SFH) of a region: $R_X[t] = \nu_X \int_{0}^t DTD[t-t'] \ SFH[t'] \ dt'$, 
where $\nu_X$ is the number of events of type $X$ that result from every solar mass of stars formed\cite{SI}.
We adopt a DTD of the form\cite{Childress2014}: $DTD[t] \propto \frac{(t/t_p)^\alpha}{(t/t_p)^{\alpha-s}+1}$, 
where $t_p$ is the characteristic `delay time'.
For the SN~Ia DTD, we set $t_p = 0.3$ Gyr, $s = -1.0$, and  $\alpha = 4$ following ref.~\cite{Childress2014}.
From fitting the BPS data on the CO-WD He WD mergers we find\cite{SI}, 
$t_p = 5.4^{+0.8}_{-0.6}$ Gyr, $s = -1.6^{+0.4}_{-0.5}$, and $\alpha = 3.7^{+1.2}_{-1.0}$ (1$\sigma$ errors).
The vertical dashed lines are at 0.3 and 5.4 Gyr.
}
\label{fig_plotDTDforPaper}
\end{figure}

However, the updated empirical situation now allows us to address the question of positron source age more quantitatively.
To do this, we use the formalism 
of Delay Time Distributions (DTDs; Fig.~\ref{fig_plotDTDforPaper}) and the known star formation histories (SFH) of the bulge, disc, and nucleus (Fig.~\ref{fig_plotSFRMW}), to find the delay time $t_p$  between star formation and the formation of 
a transient source that can reproduce the inferred\cite{Siegert2015} $B/D$ and $N/B$ positron luminosity ratios.
We adopt the parsimonious assumption 
(retrospectively justified by a global analysis of the  data\cite{SI})
that the time-integrated efficiency for positron source creation
per unit mass star formation is invariant throughout the Galaxy.

\begin{figure}%[hb!]
\centering
\includegraphics[width = 0.9 \textwidth]{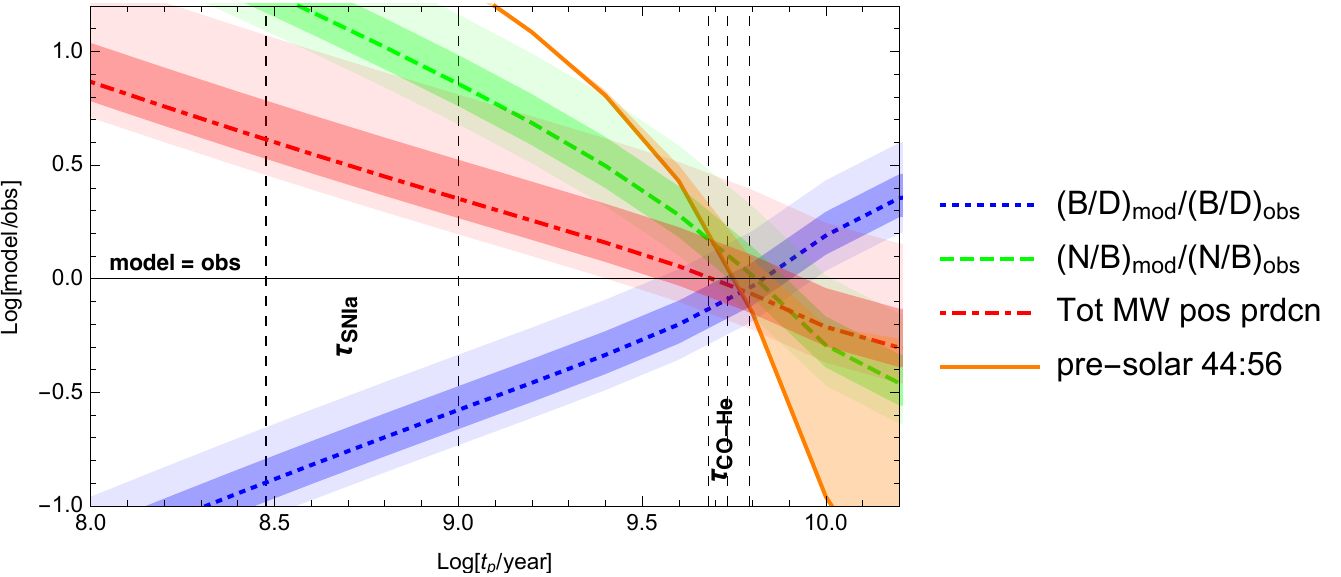}
\caption{{\small {\bf Constraints on the characteristic delay time $t_p$ of the main source of Galactic positrons.}
Curves plot the following quantities as a function of the delay time $t_p$: 
{\bf `(B/D)$_\mathrm{mod}$/(B/D)$_\mathrm{obs}$' (blue,dotted)}:- 
ratio of current, modelled bulge to disc positron source occurrence  divided by the observationally-inferred\cite{Siegert2015} ratio of bulge to disc positron luminosities $0.42 \pm 0.09$;  
{\bf `(N/B)$_\mathrm{mod}$/(N/B)$_\mathrm{obs}$' (green,dashed)}:- ratio of current, modelled nucleus to bulge positron source occurrence divided by the observationally-inferred\cite{Siegert2015} ratio of nucleus to bulge positron luminosities $(8.3 \pm 2.1) \times 10^{-2}$;
{\bf `Tot MW pos prdcn' (red,dot-dashed)}:- ratio of the modelled $^{44}$Ti
positron luminosity of the Milky Way to the observed, non-$^{26}$Al luminosity (central value $4.6 \times 10^{43}$ s$^{-1}$[\cite{Siegert2015}]) for 0.029 $\msun$ $^{44}$Ti per explosion and $f_\mathrm{91bg,bulge} = 0.32$;
{\bf `Pre-solar 44:56' (orange,solid)}:-
ratio of the mass of $^{44}$Ca to $^{56}$Fe in pre-solar material
divided by the observed\cite{Lodders2003} value $1.34 \times 10^{-3}$ (negligible error) adopting a CC and SN91bg $^{56}$Fe yield of 0.1 $\msun$ and a SN~Ia $^{56}$Fe yield of 0.6 $\msun$ and assuming i) 0.029 $\msun$ $^{44}$Ti per SN91bg (lower bound on band) and ii) 0.029 $\msun$ $^{44}$Ti per SN91bg and, in addition, $1 \times 10^{-4}$ $\msun$ $^{44}$Ti per CC and 
$3 \times 10^{-5}$ $\msun$ $^{44}$Ti per SN~Ia (upper bound on band). The heavy and light coloured bands cover the 1$\sigma$ and 2$\sigma$ uncertainties, respectively, in each
measured quantity (except for the orange band which spans  the systematic uncertainties in mean
CC and SN~Ia $^{44}$Ti production). 
The group of vertical lines labelled by $\tau_\mathrm{CO-He}$ shows the central value and $\pm 1\sigma$ extent of the best fit delay time for the CO WD - He WD mergers in the BPS model; 
the vertical lines at  300 and 1000 Myr bracket the characteristic delay time for SNe Ia (e.g. ref.~\cite{Childress2014});  these occur at too short a delay time to  explain the $e^+$ phenomenology.}
}
\label{fig_plotCombinedConstraints}
\end{figure}

The blue and green bands in Fig.~\ref{fig_plotCombinedConstraints} show 
modelled, current $B/D$ and $N/B$  normalized to the observationally-inferred positron injection ratios 
as a function of the positron source delay time $t_p$.
Given that both bands show agreement of model and observation
for $t_p \sim 3-6$ Gyr, we conclude that a single type of positron source, arising in stellar objects with this characteristic age and older, can explain the
gross distribution of positron annihilation in the Galaxy.

But what is this source?
We have already seen the positron injection energy constraints rule out many scenarios for positron creation.
However, the required low injection energies are entirely compatible with a $e^+$ origin in $\beta^+$ decay of radionuclides 
synthesised in stars and/or stellar explosions\cite{Higdon2009,Prantzos2011}.
Important, positron-producing decay chains (with their markedly different total lifetimes shown in parentheses)
are: $^{56}$Ni $\to ^{56}$Co $\to ^{56}$Fe (${\sim}80$ days), 
$^{44}$Ti $\to ^{44}$Sc $\to ^{44}$Ca (${\sim}60$ years), 
and $^{26}$Al $\to ^{26}$Mg (${\sim}717,000$ years).
We consider each of these isotopes in turn.

Given the rates of SNe~Ia and their prodigious $^{56}$Ni yield, (${\sim}0.6 \msun$ per event), 
if only a few percent\cite{Chan1993} of the positrons released in the 
$^{56}$Ni $\to ^{56}$Co $\to ^{56}$Fe decay chain could reach the ISM, they could sustain the total diffuse Galactic positron production\cite{Higdon2009}.
However, this explanation fails for two reasons: 
i) the typical delay time of SNe Ia is too short to reproduce the observed Galactic annihilation line distribution (see fig.~\ref{fig_plotCombinedConstraints} and ref.~\cite{Prantzos2011});
ii) recent analyses based on pseudo-bolometric SN~Ia light curves constructed with the inclusion of infrared data (missing in earlier studies) 
indicate complete e$^+$ trapping in SN~Ia ejecta to late times, $\gsim 800$  
days[\cite{Leloudas2009,Kerzendorf2014,Graur2016}], 
implying that very few positrons, $\ll 1\%$, escape to the ISM.

Similarly, the long decay time of $^{26}$Al (comparable to the $10^5-10^6$ year positron ISM thermalization timescale\cite{Churazov2011}) 
guarantees that the flux of the 1.809 MeV $\gamma$-ray line associated to $^{26}$Al is in steady state, while the total intensity of this line
normalises total $^{26}$Al production in the Galaxy to 
only $4 \times 10^{42}$ s$^{-1}$[\cite{Siegert2015}], $\lsim 10$\% the Galactic value. 
Moreover, the $^{26}$Al emission is distributed along the Galactic plane, correlated with massive stars\cite{Bouchet2015}, and does not match the overall annihilation morphology.

This leaves $^{44}$Ti as the sole viable radionuclide positron source.
The fact that the $^{44}$Ti $\to ^{44}$Sc $\to ^{44}$Ca decay chain has a 60 year decay time %, between the ^{56}$Ni and ^{26}$Al decay chain 
opens an interesting potentiality: 
On the one hand, such a decay time
means that 
supernova ejecta have expanded to low densities before the 
daughter positrons are released and such positrons are extremely likely\cite{Chan1993} to reach the ISM. 
On the other hand, this decay time is considerably less than the positron ISM thermalisation time meaning that, depending on the recurrence time of
Galactic $^{44}$Ti sources, the total mass of $^{44}$Ti in the Galaxy need not be  in steady state (even while the ISM daughter positrons of $^{44}$Ti are).

%As for $^{26}$Al and $^{56}$Ni, 
However, any $^{44}$Ti scenario for Galactic positron production is also constrained by the following consideration:
The observed
abundance of $^{44}$Ca relative to
 $^{56}$Fe in solar system material
 indicates that
 Galactic $^{44}$Ti production at $t_\mathrm{lookback} > 4.55$ Gyr
would have generated a positron
luminosity
$(0.3 - 1.2) \times 10^{43}$ e$^{+}$ s$^{-1}$ [\cite{Chan1993,Higdon2009,Siegert2015}], 
$\lsim 1/4$ of the current value.
Thus if $^{44}$Ti is currently the source of most positrons, the Galactic $^{44}$Ti injection rate must be
significantly larger now than between the Big Bang and the formation of the Solar System 4.55 Gyr ago.
But note: this requirement is naturally satisfied for $^{44}$Ti sources that occur at the same $\sim$ few Gyr delay times required by the
Galactic distribution of positron annihilation.

A further consideration is the following: 
It is conventionally assumed that core-collapse (CC) supernovae (SNe) deliver most $^{44}$Ti in the Galaxy via alpha-rich freezeout near the mass cut between the proto-neutron star and the ejecta.
However, while evidence exists for synthesis of few $\sim  10^{-4} \msun$ of $^{44}$Ti in three specific SN remnants\cite{Troja2014},
%:the CC SN remnant, Cas A, and the SN~Ia remnant, Tycho in the Galaxy and the CC remnant SN 1987A in the LMC\cite{The2006,Troja2014}. 
the number of remnants currently emitting in $^{44}$Ti $\gamma$-ray and X-ray lines is too small to be a comfortable match to the number expected were such sources to be responsible for most Galactic $^{44}$Ca[\cite{The2006}].
Moreover, 
CC nucleosynthesis models
do not yield sufficient  $^{44}${Ti} to explain the abundance of $^{44}$Ca in mainstream, pre solar system material\cite{Timmes1996}.  %Sun\cite{Woosley1994,Timmes1996} . 
Indeed, the origin of pre-solar $^{44}$Ca, which very likely derives from $^{44}$Ti, is a significant, unresolved problem in nuclear astrophysics.

A possible resolution of all these anomalies is that there are events, rarer than
CC SNe,
 that supply significantly larger masses of $^{44}$Ti than CC SNe seem capable of\cite{Leising1994}.
To comfortably obey the observational constraints\cite{SI}, these events should 
i) currently occur in the Galaxy every $\gsim 300$ years 
(a few times longer than the decay time of the $^{44}$Ti decay chain so that, as presaged above, the mass of $^{44}$Ti in the Galaxy is not in steady state so that neither is the 
$^{44}$Ti $\gamma$-ray and X-ray line flux 
from the Galaxy steady);  
and ii) feature a minimum mean $^{44}$Ti yield:
\bea
\left<M_{^{44}Ti}\right> \gsim 0.013 \msun \left(\frac{R}{(300 \ \mathrm{year})^{-1}} \right)  \left(\frac{\dot{N}_{e^+}}{5 \times 10^{43} \ \mathrm{s}^{-1}} \right)
 \, .
 \label{eqn_Min44TiMass}
 \eea

In summary, a universal, stellar positron source in the Galaxy that explains both the morphology and total amount of positron annihilation would:
i) occur at a characteristic delay $\sim 3-6$ Gyr subsequent to star formation and be, therefore, more frequent in today's Galactic disc than at early times;
ii) have  a rate evolving according to this characteristic delay and currently synthesise $5.8^{+1.3}_{-1.9} \times 10^{-5} \msun$ year$^{-1}$ of $^{44}$Ti 
(saturating the total Galactic positron luminosity minus that due to $^{26}$Al decay);
iii) achieve this by yielding $\gsim 0.013 \msun$ of $^{44}$Ti per event and occurring with a mean repetition time $\gsim 300$ years.
Such a source would simultaneously explain the origin of pre-solar $^{44}$Ca and naturally address the lack of strong $^{44}$Ti $\gamma$-ray and X-ray line sources in the current sky.

This large $^{44}$Ti yield can probably only be satisfied by astrophysical 
helium detonations in which incomplete burning of $^{4}$He 
leads\cite{Hansen1971,Woosley1986} to nucleosynthesis products dominated by intermediate mass $\alpha$-nuclei such as $^{40}$Ca, $^{44}$Ti, and $^{48}$Cr.
For optimal conditions up to $\sim$1/3 of the $^{4}${He} can be burned to $^{44}$Ti for adiabatically compressed He-WD matter. 

\begin{figure}%[hb!]
\centering
\includegraphics[width = 1 \textwidth]{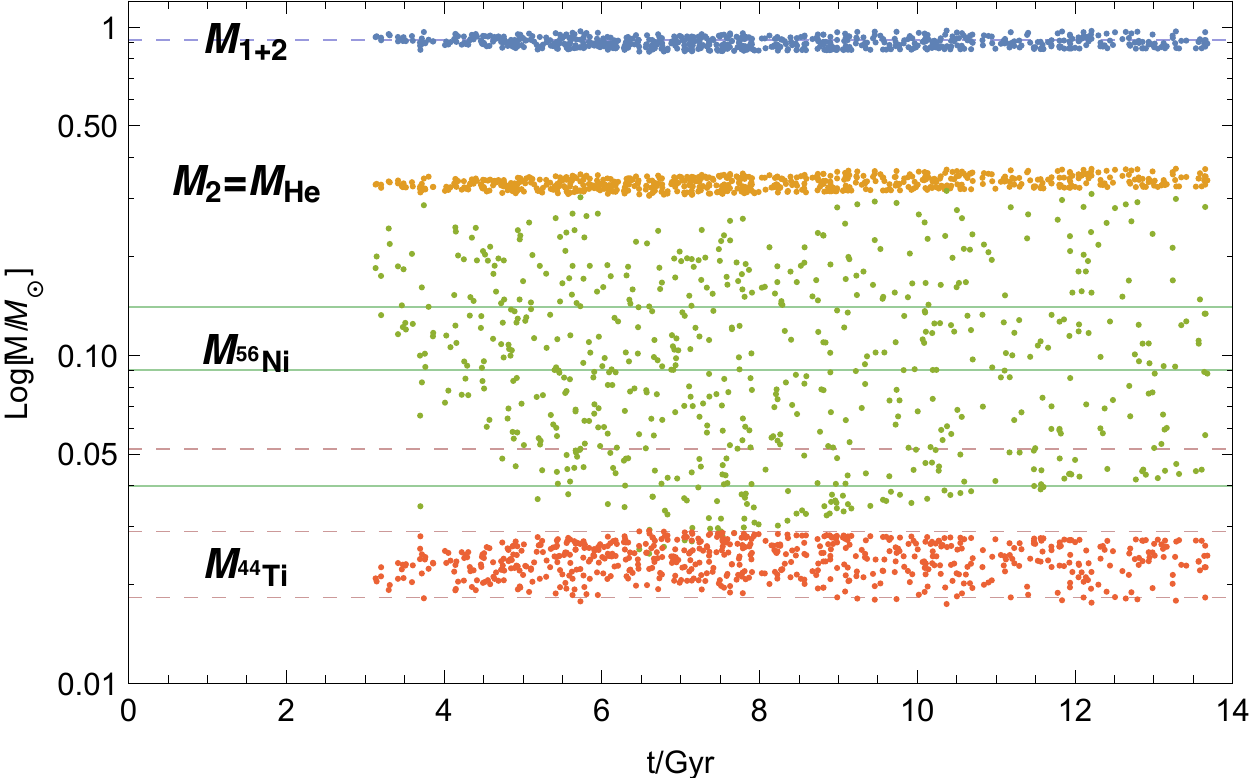}
\caption{{\bf Modelled total (blue dots) and secondary (He; yellow dots) WD masses at the time of merger and the inferred synthesised masses of $^{56}$Ni (green dots) and $^{44}$Ti (orange dots) vs. delay time.}
We assume the CO and He WDs merge to form an object (of mean mass 0.91 $\msun$: dashed blue line) that reaches quasi-hydrostatic equilibrium before exploding.
For the $^{56}$Ni  yield we
use the combined mass of both WDs to 
which we apply an interpolation\cite{Piro2014} of the modelled\cite{Sim2010} mass of $^{56}$Ni yielded in detonations of single, sub-Chandrasekhar WDs; 
the rather steep dependence of the $^{56}$Ni yield on total mass is reflected in the scatter of the green points.
The $^{56}$Ni yields (mean $0.11 \msun, \sigma = 0.05 \msun$) are
within the range of the observationally-inferred\cite{Sullivan2011}  $^{56}$Ni mass 
for a sample of 15 SNe 91bg 
(mean $0.09 \msun, \sigma = 0.05 \msun$ shown as
the solid green lines).
The $^{44}$Ti yield of this channel,  which depends on the amount of He in the correct density range ($\sim 10^5-10^6 \ \mathrm{g\,cm^{-3}}$), is similarly estimated\cite{SI}
at $(0.024\pm0.0028) \msun \ (1\sigma)$, consistent with the $0.029^{+0.18}_{-0.11} \msun$ determined in the main text for the Galactic positron source and shown as the dashed red lines.
}
\label{fig_plotBPSResults}
\end{figure}

Our binary evolution population synthesis (BPS) models\cite{Karakas2015} 
employing StarTrack\cite{Belczynski2008},
reveal an evolutionary channel  that is expected to aggregate large masses of helium at the high densities suitable for detonation at long timescales after star formation.
This channel (``channel 3" in ref.~\cite{Karakas2015}) involves low-mass ($\sim 1.4-2 \msun$), 
interacting binary star systems that 
experience two mass transfer events (with only the second being a common-envelope interaction), evolve into a CO WD and a (pure) 0.31--0.37 $\msun$ He WD (with the progenitor of the latter never undergoing helium core burning),
and subsequently merge (see Fig.~\ref{fig_plotBPSResults}). 
These mergers occur at a characteristic timescale $t_p = 5.4^{+0.8}_{-0.6} \times 10^9$ year (2$\sigma$).
Non-trivially, this is an extremely good match to the empirical constraints on the characteristic delay time of a Galactic positron source; cf. vertical dashed lines on right of 
Fig.~\ref{fig_plotCombinedConstraints}.
%
%The best-fit late-time decay rate of the mergers has a distribution that follows $\propto t^{-s}$ where $s = 1.6_{-0.5}^{+0.4}$.

It has previously been suggested\cite{Pakmor2013, Dan2015} that mergers of CO WD--He WD
binaries may be the immediate progenitors of the class of sub-luminous
thermonuclear supernova known as SN1991bg-like
(`SN91bg':\cite{Filippenko1992}). 
Indeed, 
the $^{56}$Ni yields we estimate
for
our ``channel 3'' merger events (fig.~\ref{fig_plotBPSResults}) are 
a
good match to the empirically-determined $^{56}$Ni yields of
SN1991bg-like events (but are too large to match SN2005E-like SNe, a
class previously suggested\cite{Perets2010} to supply the Galactic
positrons).
The $^{56}$Ni yields are estimated
by assuming that the merger product assumes a transient configuration of quasi-hydrostatic equilibrium 
with the He WD secondary accreted on to the CO WD primary before, first, the helium shell detonates and, second, the CO core detonates.
%

%%Modelling evidence suggests\cite{Pakmor2013, Dan2015}  
%%that mergers of CO WD--He WD binaries are the immediate progenitors of the class of sub-luminous thermonuclear supernova known as SN1991bg-like (`SN91bg'[\cite{Filippenko1992}]).
%
%%In distinction to refs.~\cite{Pakmor2013, Dan2015}, however, and inspired by ref.~\cite{vanKerkwijk2010}, we suggest that 
%%the merger product assumes a transient configuration of quasi-hydrostatic equilibrium 
%%with the He WD secondary accreted on to the CO WD primary. 
%
%%The system then suffers a detonation in the overlying helium layer/shell, which then ignites carbon in the core.
%
%%The identification of such mergers with SNe 91bg is consistent with the $^{56}$Ni 
%%yields ($\sim 0.1 \msun$) of the modelled mergers: these are a good match to the empirically-determined $^{56}$Ni yields of SNe 91bg  (fig.~\ref{fig_plotBPSResults}).
%
%%At the same time, the  $^{56}$Ni yields imply that these mergers are not SN2005E-like SNe, a class previously suggested to supply the Galactic positrons\cite{Perets2010}.

Empirically, SNe 91bg immediately match a number of our requirements:
they are quite frequent today,
representing
$\sim 15$ \% of all thermonuclear SNe amongst 
the local galaxies sampled in the volume-limited Lick Observatory Supernova Search (LOSS[\cite{Li2011a}]) of 74 SNe Ia within 80 Mpc.
They also occur in old stellar environments like elliptical galaxies\cite{Howell2001,Perets2010}.
%\cite{Howell2001,Taubenberger2008,Li2011b,Perets2010}.
%
Moreover, the cosmological rate of SNe 91bg is increasing\cite{Gonzalez-Gaitan2011}
(within large statistical uncertainties), suggesting they are governed by delay times significantly larger than CC SNe or SNe~Ia (which are becoming less frequent in today's universe) as required by our scenario.
In addition, SNe 91bg do show evidence of titanium in their spectra, as revealed by a characteristic Ti \textsc{ii} absorption 
trough in their spectra around ${\sim}4200\,${\AA} (on this basis
their potential importance for supplying Galactic $^{44}$Ti-decay
positrons was previously noted\cite{Higdon2009}).

Since SNe 91bg represent 10 out of the 31 %$0.32^{+0.15}_{-0.18}$  ($2\sigma$) 
thermonuclear supernovae in early-type galaxies in the LOSS sample\cite{Li2011a,Piro2014,SI} 
and, spectrally, the bulge is an early type galaxy (of Hubble type E0[\cite{Prantzos2011}]) we set
the current bulge SN91bg to all SN~Ia relative rate  at $f_\mathrm{SN91bg,bulge} = 0.32\pm0.16$  ($2\sigma$)[\cite{SI}].
We determine\cite{SI}  the rate of ordinary bulge SNe~Ia (excluding SN91bg) to be 
 $R_\mathrm{SNIa,bulge} = 9.8 \times 10^{-2}$/century from which $R_\mathrm{SN91bg,bulge} = 4.6_{-2.7}^{+4.4} \times 10^{-2}$/century.
Given our assumption (retrospectively justified by an analysis of the data\cite{SI}) of a universal efficiency per unit stellar mass formed for creating positron sources (here understood to be SNe 91bg), we can then also calculate the current SN91bg rate in the disc and the nuclear bulge as $(1.4 \pm0.7) \times 10^{-1}$/century and $(4.7\pm2.3) \times 10^{-3}$/century respectively, thus implying
a SN91bg Galactic recurrence time of $t_\mathrm{wait} \sim 530$ year.

Integrating over disc SN explosions
up to 4.55 Gyr ago, we find that a mean $^{44}$Ti yield $0.029^{+0.18}_{-0.11} \msun (f_\mathrm{SN91bg,bulge}/0.32)$ is required to reproduce the observed abundance  
of $^{44}$Ca relative to $^{56}$Fe (from $^{56}$Ni synthesised in SNe~Ia and CC SNe) for a characteristic delay time consistent with the other constraints (Fig.~\ref{fig_plotCombinedConstraints} orange curve).
Adopting $M_\mathrm{^{44}Ti} = 0.029 \msun$ and the Galactic SN91bg rate already determined, we predict the 
current positron injection rate shown as the red curve in Fig.~\ref{fig_plotCombinedConstraints}.
This saturates, within errors,
the absolute positron luminosity of the Galaxy (minus the positron luminosity of $4 \times 10^{42} \ e^+$ s$^{-1}$ due to $^{26}$Al decay) within 
a $t_d$ range consistent with the other constraints.

The mean $^{44}$Ti yield per SN91bg implied by this analysis, $\sim 0.03 \msun$, is close to direct estimates we can
make for CO WD + He WD mergers using our BPS data (Fig.~\ref{fig_plotBPSResults}) 
assuming a quasi-hydrostatic configuration as seems to be warranted on the basis of our procedure for calculating the $^{56}$Ni yield of the same explosions.
It is also
well within the range found for
helium detonation yields with nuclear network codes in single-zone simulations of explosive helium burning:
for ${\sim}0.3 \msun$ of He with substantial admixtures of $^{12}$C, $^{16}$O, or $^{14}$N at a fixed density of $10^6$ g/cm$^3$, temperatures $(2-4)\times 10^9$ K and using a 203-nuclide network,
ref.~\cite{ Perets2010} finds $^{44}$Ti yields ranging up to $0.12 \msun$.  
In full numerical hydrodynamical calculations of the detonation of He shells  of mass 0.15-0.3 $\msun$ atop CO WD cores of mass $0.45-0.6 \msun$,
ref.~\cite{ Waldman2011}  found  $^{44}$Ti yields $(0.58-3.1) \times 10^{-2} \msun$.
Adding N and C `pollution' into the He shell can act to significantly increase the $^{44}$Ti yield (and decrease the $^{56}$Ni yield) of the He detonation.

Our approach -- to estimate the relevant nucleosynthesis yields by
assuming the merger remnant reaches hydrostatic equilibrium before detonating -- appears viable but awaits confirmation with high resolution, 3-dimensional numerical hydrodynamical simulations post-processed with a detailed nuclear network code.
To our knowledge, no such simulations have been performed in the correct mass range to directly test our scenario and it is thus essentially an hypothesis that the CO-He WD binary mergers detonate to yield $^{44}$Ti masses in the range $0.013-0.03 \msun$; 
we commend such simulations to the community. 
The considerations presented here do not constitute a logical proof that there is a single  type of positron source in the Galaxy.
Indeed, a $\sim 10$\% contribution\cite{Siegert2015} from $^{26}$Al from massive stars is inescapable; other sources like microquasars\cite{Guessoum2006,Siegert2016} plausibly make a further contribution.
We have shown here, however, that the  Galactic annihilation morphology  admits of a single dominant positron source type which must then be connected to stellar populations 
3-6 Gyr old. 
This is a general constraint that is difficult to evade given the extreme paucity of young stars in the Galactic bulge together with the fact that all regions of the Galaxy possess old stellar populations.

%%%%%%%%%%%%%%%%%%%%%%%%%%%%%%%%%%%%%%%%%%%%%%%%%%%%%%%%%%%%%%%%%%%%%%%%%%%%%%%

%\begin{thebibliography}{99}

%%%%%%%%%%%%%%%%%%%%%%%%%%%%%%%%%%%%%%%%%%%%%%%%%%%%%%%%%%%%%%%%%%%%%%%%%%%%%%%

\begin{addendum}
% \item Put acknowledgements here.
\item[Supplementary Information] is linked to the online version of the paper at www.nature.com/nature. 

\item[Materials and Correspondence] Correspondence and requests for materials
should be addressed
to Roland Crocker (email: rcrocker@fastmail.fm). 
 
\item[Acknowledgements]
 RMC was the recipient of an Australian Research Council Future Fellowship (FT110100108). 
Parts of this research were conducted by the Australian Research Council Centre of Excellence for All-sky Astrophysics (CAASTRO), through project number CE110001020.
DMN is supported by the Allan C. and Dorothy H. Davis Fellowship.
The authors thank  Janaina Avila, John Beacom, Nicole Bell, Geoff Bicknell, Donald Clayton, Ken Freeman, Ortwin Gerhard, Jarrod Hurley, Trevor Ireland, Amanda Karakas, Matthew Kerr, Joshua Machacek, Fulvio Melia,  Daniel Murtagh, Ryan O'Leary,
R{\" u}diger Pakmor, Thomas Siegert, Patrick Tisserand,
Ray Volkas,  
Anton Wallner, Rosemary Wyse, and Fang Yuan for very useful discussion.
They particularly thank Brian Schmidt for pointing out the potential importance of SN1991bg-like SNe to the positron problem.

\item[Author Contributions] 
All the authors discussed the results and commented on the manuscript. 
RMC wrote the paper.  
AJR performed BPS modelling used in the paper and provided theoretical input.
IRS provided theoretical input, helped with calculating yields of the Helium detonations, and contributed to the writing of the paper. 
FHP, AM, and BET provided advice about rates, prevalence, and distribution of 91bg in supernova searches
HB, LF, and JJE,  provided advice on the BPS.
AM and MW provided statistical analysis.
DMN provided advice about the SFH of the Galactic bulge and other theoretical input.
SS provided input on the phenomenology of SN explosions.
FA provided input on the phenomenology of positron transport and annihilation radiation.
All the authors commented on the draft text.

 \item[Competing Interests] The authors declare that they have no
competing financial interests. 
%Reprints and permissions information is available at npg.nature.com/reprintsandpermissions.
 %\item[Correspondence] 
\end{addendum}

%%%%%%%%%%%%%%%%%%%%%%%%%%%%%%%%%%%%%%%%%%%%%%%%%%%%%%%%%%%%%%%%%%%%%%%%%%%%%%%
\clearpage

\clearpage

%%%%%%%%%%%%%%%%%%%%%%%%%%%%%%%%%%%%%%%%%%%%%%%%%%%%%%%%%%%%%%%%%%%%%%%%%%%%%%%

\clearpage

%%%%%%%%%%%%%%%%%%%%%%%%%%%%%%%%%%%%%%%%%%%%%%%%%%%%%%%%%%%%%%%%%%%%%%%%%%%%%%%

\clearpage

%%%%%%%%%%%%%%%%%%%%%%%%%%%%%%%%%%%%%%%%%%%%%%%%%%%%%%%%%%%%%%%%%%%%%%%%%%%%%%%

\clearpage

%%%%%%%%%%%%%%%%%%%%%%%%%%%%%%%%%%%%%%%%%%%%%%%%%%%%%%%%%%%%%%%%%%%%%%%%%%%%%%%

\section*{Methods}

\subsection*{Star formation rate parameterisation}

For the disc and bulge star formation rates, we use the form suggested by ref.~\cite{vanDokkum2013}:
\begin{equation}
\log_{10}[SFR + D] = \max[A z^2 + B z + C, 0] \, .
\label{eq_SFH}
\end{equation}
For the disc the coefficients $A,B,C$, and $D$ are determined by fitting to the Milky Way disc star 
formation history data of ref.~\cite{Snaith2014} renormalized so that the integrated stellar mass of the disc (outside the VVV volume) is $(3.7 \pm 0.5) \times 10^{10} \msun$[\cite{Bland-Hawthorn2016}]; we find $\{A = -4.06 \times 10^{-2}, B = 0.331, C = 0.338, D = 0.771 \}$ (note that we account for a factor 0.26 of the star formation rate  lost to stellar winds and stellar ejecta: ref.~\cite{vanDokkum2013}).
This parameterisation gives a present-day star formation rate in the disc of 1.4 $\msun$ year$^{-1}$, in tolerable agreement with estimates in the literature (e.g., $1.65 \pm 0.19 \msun$ year$^{-1}$ for a Kroupa IMF: \cite{Licquia2015}).
For the current bulge region mass we require that the star formation rate, with form given by eq.~\ref{eq_SFH}, integrates to $1.55 \times 10^{10} \msun$[\cite{Bland-Hawthorn2016}], 
with the SFR peaking at a look-back time of 13 Gyr and going to zero by 10 Gyr[\cite{Nataf2015}].
With these constraints, the SFR can again be represented in the form of Eq.~\ref{eq_SFH} with 
$\{A = -2.62 \times 10^{-2}, B = 0.384, C = -8.42 \times 10^{-2}, D = 3.254 \}$.
The bulge and disc star formation histories are plotted in Fig.~\ref{fig_plotSFRMW} of the main text.

For the SFH of the nuclear bulge, the distinct population of stars that dominates the stellar density within 200-400 pc of the central supermassive black hole, 
there is intermittency and burstiness on various timescales (e.g., ref.~\cite{Krumholz2016}).
However,  averaging over $\gsim 100$ Myr, as pertinent to the calculation of the SN91bg rate, the nuclear star formation seems remarkably stable\cite{Figer2004}.
We therefore assume it to be constant at a rate equal to $0.14 \msun$ year$^{-1}$ that will supply the mass of the nuclear bulge in the time since the big bang.

\subsection*{Delay time distribution}

Our adopted DTD rises as  $\propto t^\alpha$ for $t \ll t_p$ and asymptotes to $\propto t^s$ for $t \gg t_p$ with the characteristic timescale $t_p$ labelled `the delay time'. 
A population whose mergers are governed purely by gravitational radiation typically has $s = -1$ \cite{Ruiter2009}; if there are other processes hastening the process of inspiral, the delay time distribution is steepened leading to $s < -1$.  
Following ref.~\cite{Childress2014} we set $\alpha = 4$, $s = -1$, and $t_p = 0.3$ Gyr for SNe~Ia (for these $\alpha$ and $s$ values, 
the maximum rate of SNe~Ia occurs at $1.32 \ t_p$ subsequent to a star formation burst).

For the SN91bg DTD we set fiducial values of $s = -1.6$ and $\alpha = 3.7$ on the basis of fitting the data for CO WD -- He WD mergers in our BPS (see fig.~\ref{fig_plotTrianglePlots} in the Supplementary Information).
Using a Markov chain Monte Carlo (MCMC) approach we fit (using the emcee package\cite{Foreman-Mackey2013}) the DTD functional form from ref.~\cite{Childress2014} $DTD[t] \propto \frac{(t/t_p)^\alpha}{(t/t_p)^{\alpha-s}+1}$ 
to the histogrammed BPS data.
From this we obtain $t_p = (5.4_{-0.6}^{+0.8}) \times 10^9$ year, $s = -1.6_{-0.5}^{+0.4}$, and $\alpha = 3.7_{-1.0}^{+1.3}$ (1$\sigma$ errors).
Note that in fig.~\ref{fig_plotCombinedConstraints} we  set $s = -1.6$ and $\alpha = 3.7$ but that $t_p$ is a free parameter we scan over; we have checked that 
setting $s = -1.0$ and $\alpha = 4.0$ we obtain  similar constraints.

\subsection*{Estimated $^{44}$Ti and $^{56}$Ni yields of He WD-CO WD mergers}

In the main text and in the caption to Figure \ref{fig_plotBPSResults}
we mention our estimates of the $^{44}$Ti and $^{56}$Ni yields of the
He WD-CO WD mergers modelled within our BPS calculation: $\sim 0.03
\msun$ and $\sim 0.1 \msun$ respectively.

To obtain these yields, we have assumed that the merger product
undergoes detonation.
The efficiency with which the explosive nuclear burning converts initial He or CO into $^{44}$Ti  or $^{56}$Ni is a function of density, temperature, 
and is also sensitive (in the case of He burning through to intermediate mass elements) to pollution of the He by, e.g., C and/or N \cite{Waldman2011,Holcomb2013}.
Thus in order to estimate the yield of $^{44}$Ti  or $^{56}$Ni by the CO -- He WD mergers we need to know 
i) the density profile of the burning material and ii) the fractional yield (of 
$^{44}$Ti  or $^{56}$Ni) as a function of density (and temperature).
In the relevant few $\times 10^9$ K temperature range, efficient conversion of $^4$He to intermediate mass elements (including $^{44}$Ti)  occurs when the $^4$He is burned at a density between $\sim 10^5$ and $10^6 \
\mathrm{g\,cm^{-3}}$[e.g., ref.~\cite{Holcomb2013}]. 
Burning of CO through to $^{56}$Ni requires that the CO be at densities above ${\sim}10^7\,\mathrm{g\,cm^{-3}}$ (we find that only a small contribution, $< 10\%$, is made to total $^{56}$Ni  from incomplete Si-burning and He-burning at lower densities).

In order to estimate the initial density profile of  He (and CO) before it detonates, we  assume that the merger
remnant temporarily assumes a configuration of quasi-hydrostatic equilibrium 
with a total mass given by the combined mass of the pre-merger primary and secondary, with the disrupted secondary's  He accreted into a shell atop the primary CO.
This sort of approach is consonant with the suggestion from
van Kerkwijk et al.~\cite{vanKerkwijk2010}
that relatively low-mass CO+CO WD mergers 
first produce a 
merged remnant 
containing most of the aggregated mass of the binary
before subsequently detonating.
On the other hand, it
is in tension with the pictures developed by Pakmor et
al.\cite{Pakmor2013} and Dan et al.\cite{Dan2015} based on their
hydrodynamical merger simulations. 
In particular, the prompt
detonation scenario\cite{Pakmor2013} does not allow
the assembly of $\gsim 0.1 \msun$ mass of $^4$He in the correct density range for $^{44}$Ti synthesis
because detonation is triggered
via He ignition shortly after mass transfer begins.
Dan et
al.\cite{Dan2015} investigate the possibility of explosions at a later
phase, shortly after the dynamical merger is complete. 
Of their models,
they suggest a best match with SN1991bg-like explosions from the
simulated merger of a 0.45 $\msun$ He WD with a 0.9 $\msun$ CO
WD. This system is significantly more massive than the population of
mergers identified in our BPS calcualtions and, moreover, they find
that this model would yield only $\sim 3 \times 10^{-5} \msun$ of
$^{44}$Ti, insufficient for our scenario.
Thus 
we require that explosions occur in less massive mergers
(and likely at later phases) than explored by, e.g., Pakmor et
al.\cite{Pakmor2013} and Dan et al.\cite{Dan2015}.

We gain some confidence for this picture 
%On the other hand, in support of the hydrostatic picture we adopt, the work 
%of van Kerkwijk et al.~\cite{vanKerkwijk2010}
%for relatively low-mass CO+CO WD mergers does suggest that a 
%merger remnant containing most of the aggregated mass of the binary is %fully formed before subsequently detonating.
%
by considering, first, $^{56}$Ni yields from the mergers which we calculate in two ways.
First, we produce an interpolation of the fractional $^{56}$Ni yield of CO burning as a function initial density as presented in fig.~A1 of ref.\cite{Fink2010}.
We then apply this parameterized yield to CO+He merger remnant density profiles generated by assuming that the remnants reach hydrostatic equilibrium before detonating.
Second, we use an existing interpolation\cite{Piro2014} of the modelled\cite{Sim2010} yield of $^{56}$Ni in detonations of single, sub-Chandrasekhar WDs that is given as a function purely of the total mass of the WD.
We find excellent agreement between these two approaches (see fig.S5 in the Supplementary Information\cite{SI}).
We then observe that, for our modelled distribution of CO+He WD mergers,
applying either of these approaches, we derive $^{56}$Ni yields that
are very close to the observationally-inferred $^{56}$Ni
yields of the sample of SNe 91bg assembled in ref.~\cite{Sullivan2011}
(see caption to fig.~\ref{fig_plotBPSResults}); this is an independent consistency check of our model.

We calculate $^{44}$Ti yields from the He detonation in the merged He+CO remnants in a similar fashion to $^{56}$Ni:
i) the initial He density profile is obtained assuming the merger remnant achieves hydrostatic equilibrium before detonation; 
ii) this density distribution is convolved with the fractional yield of $^{44}$Ti from He detonation given as a function of density  
in ref.~\cite{Holcomb2013} (using our own interpolation of their results and adopting a  fixed temperature
of $2 \times 10^9$ K (e.g., \cite{Moore2013}).

\renewcommand\refname{Methods Section References}

\noindent
{\bf Data availability:} The data that support the plots within this paper and other findings of this study are available from the corresponding author upon reasonable request.

\newpage

%%%%%%%%%%%%%%%%%%%%%%%%%%%%%%%%%%%%%%%%%%%%%%%%%%%%%%%%%%%%%%%%%%%%%%%%%%%%%%%

\section*{Supplementary Information to: Diffuse Galactic Antimatter from Faint Thermonuclear Supernovae  in Old Stellar Populations}
\setcounter{section}{1}

\noindent
Roland~M.~Crocker$^{1}$, %\thanks{E-mail: rcrocker@fastmail.fm}\thanks{Australian Research Council Future Fellow}, 
Ashley~J.~Ruiter$^{1,2}$,
Ivo~R.~Seitenzahl$^{1,2,3}$,
Fiona~H.~Panther$^{1,2}$,
Stuart Sim$^{4}$,
Holger Baumgardt$^{5}$,
Ana{\" i}s M{\"o}ller$^{1,2}$,
David~M.~Nataf$^{1,6}$,
Lilia Ferrario$^{7}$,
%Jarrod Hurley${^5}$,
J.J. Eldridge$^{8}$,
Martin White$^{9}$,
Brad~E.~Tucker$^{1,2}$,
and Felix Aharonian$^{10,11}$

\begin{affiliations}
 \item Research School of Astronomy and Astrophysics, Australian National University, Canberra, Australia
  \item ARC Centre of Excellence for All-sky Astrophysics (CAASTRO), Canberra, Australia
 \item School of Physical, Environmental and Mathematical Sciences, UNSW Canberra, Australian Defence Force Academy, Canberra, Australia
 \item School of Mathematics and Physics, Queen's University, Belfast, U.K.
\item School of Mathematics and Physics, University of Queensland, Brisbane, Australia
\item Department of Physics and Astronomy,
The Johns Hopkins University, Baltimore, U.S.A.
 \item Mathematical Sciences Institute, Australian National University, Canberra, Australia
 %\item Centre for Astrophysics and Supercomputing, Swinburne University of Technology, Melbourne, Australia
 \item Department of Physics, University of Auckland, Auckland, New Zealand
\item Department of Physics, University of Adelaide, Adelaide, Australia
\item Dublin Institute for Advanced Studies, Dublin 2, Ireland
\item Max-Planck-Institut f{\" u}r Kernphsik, Heidelberg, Germany
\end{affiliations}

%%%%%%%%%%%%%%%%%%%%%%%%%%%%%%%%%%%%%%%%%%%%%%%%%%%%%%%%%%%%%%%%%%%%%%%%%%%%%%%

\subsection{Galactic positron annihilation phenomenology in more detail}

The Galactic positron annihilation signal can be decomposed into\cite{Skinner2014,Siegert2015,Bouchet2010}: 
\begin{enumerate}
\item
A disc-like component tracing the Galactic plane with Gaussian longitude width of $\sigma_l \simeq 60^\circ$ and latitude width  $\sigma_b \simeq 10^\circ$; 
this component's total 511 keV line flux of $1.66 \pm 0.35 (1\sigma) \times 10^{-3}$ cm$^{-2}$ s$^{-1}$ together with a characteristic distance to the annihilations of 10 kpc[\cite{Siegert2015}] implies a  disc positron annihilation rate of $\sim 3 \times 10^{43}$ s$^{-1}$ (equal to the disc positron luminosity in steady state);
\item
A component coincident with the Galactic stellar bulge, in turn
composed of two\cite{Bouchet2010,Skinner2014,Siegert2015} radially-symmetric Gaussians: i) a `narrow bulge' component
centred at $(l, b) = (-1.25^\circ , -0.25^\circ )$ with Gaussian width of $\sigma_l = \sigma_b = 2.5^\circ$; ii) a `broad bulge' component 
centred at $(l, b) = (0^\circ ,0^\circ )$  extending to 
 $\sigma_l = \sigma_b = 8.7^\circ$.
 The bulge component's total 511 keV line flux of  $9.6 \pm 0.7 (1\sigma) \times  10^{-4}$ cm$^{-2}$ s$^{-1}$ together with a characteristic bulge distance of 8.5 kpc[\cite{Siegert2015}] implies a  bulge positron annihilation rate of $\sim 2 \times 10^{43}$ s$^{-1}$.
\item
A third, weak ($\sim 10^{42}$ s$^{-1}$) but distinct  annihilation component of $\sim 2.5^\circ$ in size coincident with the Galactic nucleus whose existence has only just been demonstrated\cite{Skinner2014,Siegert2015} at better than 5$\sigma$ significance. 
\end{enumerate}
Note that positron luminosities ascribable to the various Galactic structures are model-dependent to the extent that they assume the particular mean distances to the annihilation sites 
within these structures given above (see below).

Spectrally, the overall Galactic annihilation line signal may be decomposed into broad and narrow gaussians to which may be attributed $\sim 1/3$ and $\sim 2/3$ the total line flux, 
respectively\cite{ Prantzos2011}.
At present, statistical uncertainties prevent very firm conclusions on the question of whether the 511 keV annihilation signals detected from different regions of the Galaxy exhibit different spectral widths.
The observed dominance of the narrow component is consistent with a scenario in which most positrons annihilate in a warm, thermally-unstable medium of $\sim 10^4-10^5$ K[\cite{ Churazov2011}] though substantial annihilation fractions in dust or cold gas phases are not excluded\cite{Guessoum2010}.
Continuum $\gamma$-ray measurements below 511 keV indicate that $97\pm2$\% of the $e^+$ annihilation occurs through the formation of positronium.

\subsection{Compatibility between various definitions of `disc' and `bulge'}

An important point is that the spatial correspondence between the distributions of the annihilation radiation and the positron sources likely breaks down on scales $\sim$ few $\times$ 100 pc:
Positrons must lose energy before themalizing and they may travel distances approaching $\sim$ few $\times$ 100 pc while doing so\cite{Martin2012,Alexis2014}; they must also find electron partners with which to finally annihilate.
Thus, the detailed volumetric emissivity of the annihilation radiation is,  in principle, a complicated convolution  of the distribution of positron injectors, spatially-dependent transport processes, and the target gas distribution (that provides the electron annihilation partners).
On the other hand, on scales $\gsim$ kpc the gross spatial correlation between the injection distribution and the annihilation radiation distribution will be preserved; we exploit this expectation in this work.

In the main text we do not address the correspondence between the distribution of the annihilation radiation and the underlying stellar distributions of
the various named Galactic structures.
In fact, the three separate components into which Siegert et al.\cite{Siegert2015} resolve the annihilation radiation sky are geometrically defined and not related in any {\it a priori} way to stellar structures (or radiation at any other wavelength).
Nevertheless, as we now explain, these geometrically-defined annihilation radiation components do mirror Galactic stellar structures quite well.

The review and compilation of the literature presented in ref.~\cite{Bland-Hawthorn2016}   determines that the three dimensional structure of the Galactic disc
may be well represented as
 composed of thin and thick components.
 The former has an exponential scale height of $300\pm50$ pc and a scale length of $2.5\pm0.4$ kpc.
The latter, dominated by systematically older stars, has a scale height of $900\pm100$\,pc and a scale length of $2.0\pm0.2$ kpc.
Siegert et al.\cite{Siegert2015} estimate from the best-fit latitude extent for their empirical disc model  a rather large physical scale height $\sim 1$ kpc for the annihilation radiation;
this is consistent with the  scale height of the Galactic thick disc and again supports\cite{Bouchet2010} the notion that the positron sources occur among the old stars.

The definition of the bulge given in ref.~\cite{Bland-Hawthorn2016} is considerably more geometrically complicated than the disc.
It contains a significant contribution from the Galactic bar which is inclined at a small angle, $\sim 30^\circ$, to (Galactic east of) the line of sight to the Galactic Centre.
The modern understanding is that the bulge of the Milky Way is dominantly composed of stars folded up out of the disc -- by the same instability that formed the bar  --
into a `box/peanut' structure.
The `classical' component of the bulge, formed by mergers early in the history of the Galaxy, is now known to be a small, potentially vanishing, component of the bulge.

The `bulge' mass quoted to in the main text, $(1.55\pm0.15) \times 10^{10} \msun$ refers to the total stellar mass (including disc stars) within the volume of VVV survey\cite{Minniti2010} 
which is defined by
$(x \times y \times z) = (\pm 2.2 \times \pm 1.4 \times \pm 1.2)$ kpc$^3$ where the $x$ coordinate extends along the Galactic bar and the coordinate system origin is at the Galactic Centre.
The solid angle subtended by this volume is approximated by $(l \times b) \simeq (\pm 10^\circ \times \pm 10^\circ)$.
Inside this solid angle, the proportion of the total flux (integrated over the sky) from the `disc' source fitted to the annihilation data by Siegert et al.~\cite{Siegert2015} is 0.15.
This is an excellent match to the fraction of the total mass of the 3D disc stellar mass model from ref.~\cite{Bland-Hawthorn2016} that is inside the same $(\pm 10^\circ \times \pm 10^\circ)$ 
solid angle but outside the VVV volume (i.e., in the foreground and background to the VVV volume) -- also 0.15.
Thus, demonstrating the gross consistency between the different 
definitions of `disc' and `bulge', we have that the component of the total disc annihilation emission that occurs within the solid angle of the  
$(\pm 10^\circ \times \pm 10^\circ)$ bulge field but outside the VVV bulge volume is contributed by the stars that are in front of and behind the bulge in projection.
The integrated stellar mass within this solid angle {\it and} inside the VVV volume, we call the total bulge mass (irrespective of whether the stars are `true' bulge stars or not); 
this mass is\cite{Bland-Hawthorn2016}  $(1.4-1.7)  \times 10^{10} \msun$ (which we treat as $(1.55\pm0.15) \times 10^{10} \msun$).

\subsubsection{Mean distance to `disc' and `bulge' volumes}

As mentioned above, in going from a total annihilation radiation flux integrated across the sky to a positron annihilation rate, a mean distance to the annihilation region must be assumed.
Siegert et al.~\cite{Siegert2015} adopt effective distances to the bulge of 8.5 kpc, and to the disc of 10.0 kpc.
From the stellar mass model from ref.~\cite{Bland-Hawthorn2016} we find
density-weighted root mean square distances to the disc (outside the VVV bulge volume) of 10.3 kpc and to the bulge VVV volume of 8.3 kpc; 
this confirms the broad compatibility between the geometric model implicit to the Siegert et al.~\cite{Siegert2015} results and the 3D stellar distributions we assume.

\subsection{Number of supernovae arising  per unit mass of stars formed}

The caption to the main text Fig.~\ref{fig_plotDTDforPaper} contains
$\nu_X$, which is the number of supernovae of type $X$ that result from every solar mass of stars formed.
This is given by
\begin{equation}
\nu_X = \eta_X \frac{\int_{M_\mathrm{min,X}}^{M_\mathrm{max,X}}  \Psi[M] \ dM}{\int M \Psi[M] \ dM} \ ,
\end{equation}
where 
$\Psi[M]$ is stellar initial mass function, the integral in the denominator is over the entire stellar mass distribution, 
and
$\eta_X$ is the efficiency factor for the conversion of stars in the mass range 
$M_\mathrm{min,X} \to M_\mathrm{max,X}$
into type $X$ supernova progenitors (where $M_\mathrm{min,X}$ and $M_\mathrm{max,X}$ are the minimum and maximum zero age main sequence stellar masses, respectively, for progenitors of  type $X$ supernovae).

\subsection{Fitted delay time distribution of CO WD -- He WD mergers}

\begin{figure}[hb!]
\renewcommand{\thefigure}{S1}
\centering
\includegraphics[width = 1 \textwidth]{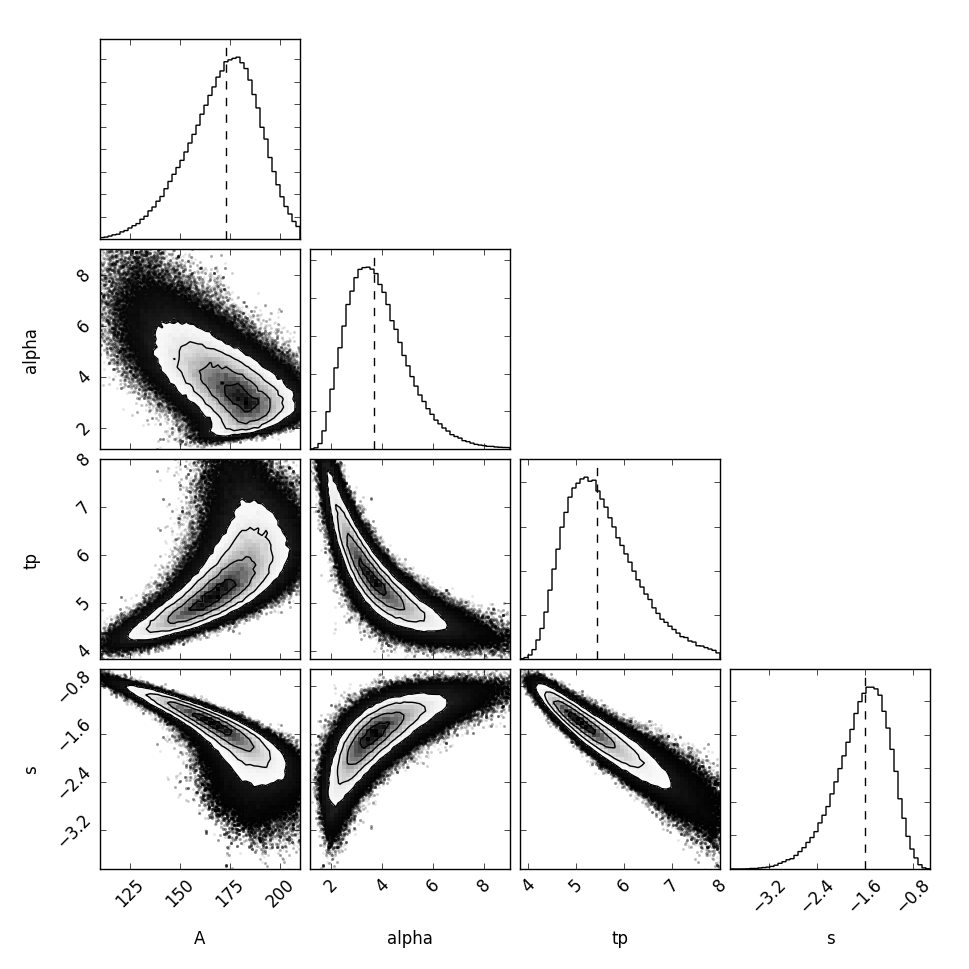}
\caption{{\bf Triangle plots for the CO WD -- He WD merger channel in our BPS}; 
$A$ is the normalization parameter.
}
\label{fig_plotTrianglePlots}
\end{figure}

Supplementary Fig.~\ref{fig_plotTrianglePlots} shows the  DTD parameters fitted (using the emcee package\cite{Foreman-Mackey2013}) to our BPS data on CO WD -- He WD mergers.

\subsection{(In)sensitivity to result of plausible variation to the assumed nuclear star formation history}

We have also rederived the $N/B$ curve under the assumption that the nuclear star formation rate parallels that of the disc (with suitably renormalized amplitude); this has a minor effect (see Supplementary Fig.~\ref{fig_nuclearSFHwithVariantSFH}), indicating our conclusions are not sensitive to systematic uncertainties in the nuclear star formation history.

\begin{figure}%[hb!]
\renewcommand{\thefigure}{S2}
\centering
\includegraphics[width = 1 \textwidth]{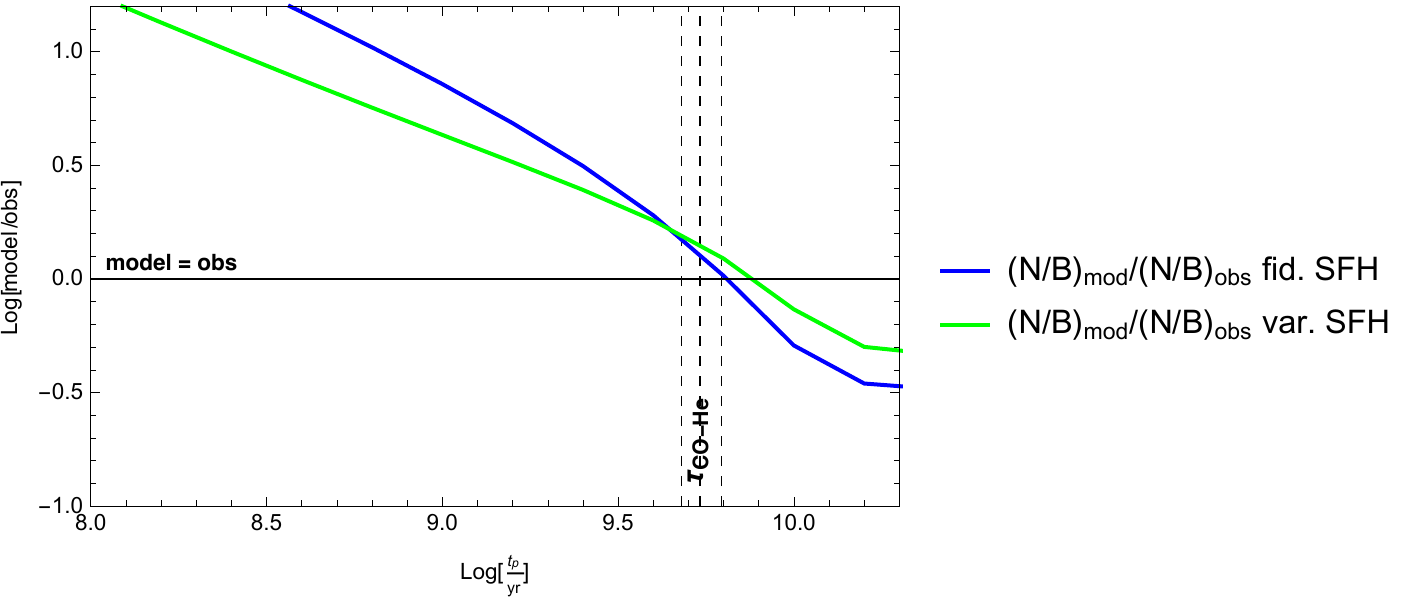}
\caption{{\bf Ratio between the nuclear and the bulge positron injection event rates for different nuclear star formation histories:}
i) a fiducial, constant star formation rate (`fid') of  $0.14 \msun$ year$^{-1}$ and ii) a nuclear star formation rate that is a fixed proportion (5\%) of the disc star formation rate; we consider that these two choices span the reasonable uncertainties in the nuclear star formation histories.
}
\label{fig_nuclearSFHwithVariantSFH}
\end{figure}

\subsection{Constraints on the minimum $^{44}$Ti yield per event and the expected wait time between events}

For putative $^{44}$Ti-yielding events 
to supply the Galactic positrons in steady state today, they
would have to occur more frequently than the positron thermalisation and annihilation timescale in the ISM of $\sim 10^5-10^6$ year[\cite{Churazov2011}].
On the other hand,
to avoid a fine-tuning to explain the fact that we do not  currently detect a strong $^{44}$Ti source via the $^{44}$Sc daughter's 67.87 keV, 78.32 keV, or 1.157 MeV $\gamma$-ray emission,
the expected wait time between $^{44}$Ti injection events must be $\gsim$ few $\times \tau_{44} \sim 300$ year, as we now show.

There are two constraints on the combination of the 
$^{44}$Ti yield per event  M$_{44}$ and the expected wait time between events $t_\mathrm{wait}$:
i) to saturate the central value of the current positron annihilation rate in the Galaxy, $5 \times 10^{43}$ e$^+$ s$^{-1}$, the time-averaged $^{44}$Ti mass injection rate
satisfies
\begin{equation}
\dot{M}_{44} = 6.3^{+1.3}_{-1.9} \times 10^{-5} \msun/\mathrm{year} \, ;
\end{equation}
this relation is plotted as the blue region in Supplementary Fig.~\ref{fig_plotRoughLowerMassLimit};
ii) the fact that we do not see a strong $^{44}$Ti $\gamma$-ray line source in the sky should not be extraordinarily fine-tuned.
A full quantitative evaluation of this constraint requires a Monte Carlo \cite{The2006,Dufour2013} but also introduces a model-dependence.
The current brightness threshold for detection of 1.157 MeV $^{44}$Ti $\gamma$-ray line sources is $\sim 1 \times 10^{-5} \ \gamma$/cm$^2$ s$^{-1}$[\cite{ The2006}]
on the basis of COMPTEL observations.
For a distance of 8 kpc (rough mean distance to the Galactic bulge), this threshold corresponds to a $^{44}$Ti mass of $3.8 \times 10^{-4} \msun$. 
If we demand that the expected current mass of $^{44}$Ti left, after decays, in the last SN91bg to have exploded in the Galaxy is less than this amount, we can derive the following constraint:
\begin{equation}
t_\mathrm{wait} \gsim \frac{\tau_{1/2}}{\ln[1/2]} \ln\left[\frac{3.8 \times 10^{-4} \msun}{\mathrm{M}_{^{44} \mathrm{Ti}}} \right] \simeq \, 90 \ \mathrm{year} \times \ln\left[\frac{3.8 \times 10^{-4} \msun}{\mathrm{M}_{^{44} \mathrm{Ti}}} \right],
\end{equation}
where $\tau_{1/2} = 60.0$ year is the  $^{44}$Ti half life.
Note this relation assumes that the expected observation time approaches $t_\mathrm{wait}$ after the last 91bg event; it is plotted as the orange region in 
Supplementary Fig.~\ref{fig_plotRoughLowerMassLimit}

\begin{figure}%%[hb!]
\renewcommand{\thefigure}{S3}
\centering
\includegraphics[width = 1 \textwidth]{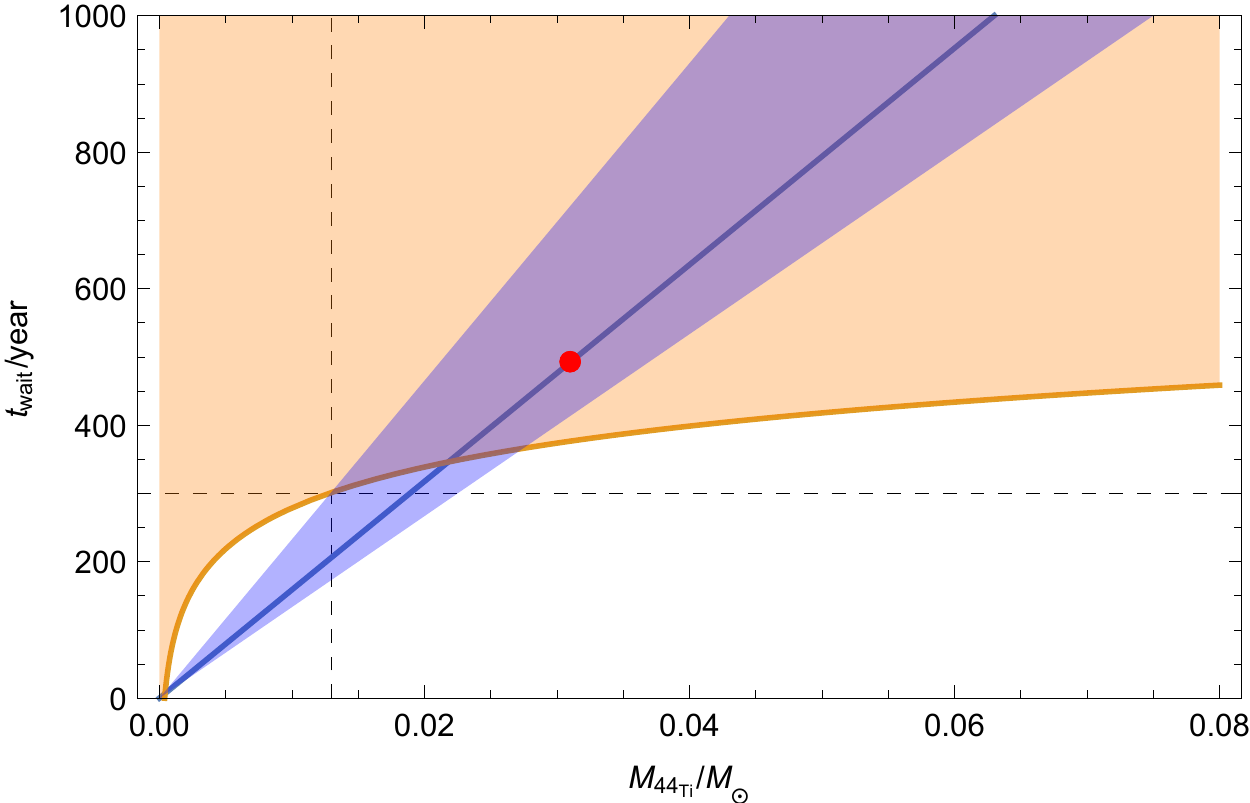}
\caption{{\bf Approximate constraints on the combination of the expected wait time between events, $t_\mathrm{wait}$, and the $^{44}$Ti yield per event, M$_{44}$}.
The allowed region is the intersection of the blue and orange zones; evidently M$_{^{44} \mathrm{Ti}} \gsim 0.013 \msun$ (vertical dashed line) and  $t_\mathrm{wait} \gsim 300$ year (horizontal dashed line).
The red dot shows the fiducial parameters derived in the main article. 
Note that $^{44}$Ti yields larger than $\sim 0.1 \msun$ are implausible for any scenario of astrophysical He detonation.
}
\label{fig_plotRoughLowerMassLimit}
\end{figure}

\subsection{Absolute rate of SNe 91bg in the bulge, disc, and nucleus}

To determine the absolute rate of
SN91bg in the Galaxy we use the following procedure. 
First we adopt a current SN~Ia rate (excluding SNe 91bg) in the disc of $R_\mathrm{SNIa,disc} = 0.42$/century[\cite{Prantzos2011}].
This means that, for our fiducial SN~Ia $t_d = 0.3$ Gyr[\cite{Childress2014}], 
$\nu_\mathrm{SNIa} = 1.96 \times 10^{-3}/\msun$
(see caption to the main text Fig.~\ref{fig_plotDTDforPaper}).
This implies a current bulge SN~Ia rate (excluding SNe 91bg) of $R_\mathrm{SNIa,bulge} = 9.8 \times 10^{-2}$/century and a nuclear bulge SN~Ia rate $R_\mathrm{SNIa,NB} = 3.1 \times 10^{-2}$/century or a
ratio  (bulge + nuclear bulge)/disc for SN~Ia rates of $(R_\mathrm{SNIa,bulge}+R_\mathrm{SNIa,NB})/R_\mathrm{SNIa,disc} = 0.29$,  in the middle of the range, 0.21 - 0.43, determined in ref.\cite{ Prantzos2011}.
Next, we use the fact\cite{Piro2014} that, of
the 31 thermonuclear supernovae in early-type galaxies in the LOSS sample, 10 are SN91bg-like, implying a 
relative rate of $f_\mathrm{SN91bg,early} = 0.32\pm0.16 \ (2\sigma)$.
Now we use the determination that the bulge is of spectral type E0[\cite{Prantzos2011}]
to set $f_\mathrm{SN91bg,bulge} = f_\mathrm{SN91bg,early}$.
This normalises the central value for the current absolute SN91bg rate at $0.32\pm0.16$ of the total bulge thermonuclear rate,
$R_\mathrm{therm,bulge} = R_\mathrm{SNIa,bulge} + R_\mathrm{SN91bg,bulge}$ giving $R_\mathrm{SN91bg,bulge} = 4.6_{-3.0}^{+4.1} \times 10^{-2}$ century$^{-1}$.
With this formalism we can also calculate that the central value of the
current disc SN91bg rate relative to all SNe~Ia is  $f_\mathrm{SN91bg,disc} = 0.18$.

We can compare these rates with those found for the CO WD - (pure) He WD merger channel in our binary population synthesis\cite{Karakas2015}.
Calculating the absolute rates for this channel (at the current age of the universe) by normalizing by stellar mass for the mass of the disc, bulge, and nuclear bulge, combined with their respective star formation histories, we find an absolute Galactic CO WD - He WD rate of $7.9 \times 10^{-4}$ year$^{-1}$, or 0.55 of the rate determined above, which we regard as excellent agreement.

\subsection{Evidence that star formation across the Galaxy is equally efficient at producing positron sources}

\begin{figure}%[hb!]
\renewcommand{\thefigure}{S4}
\centering
\includegraphics[width = 1 \textwidth]{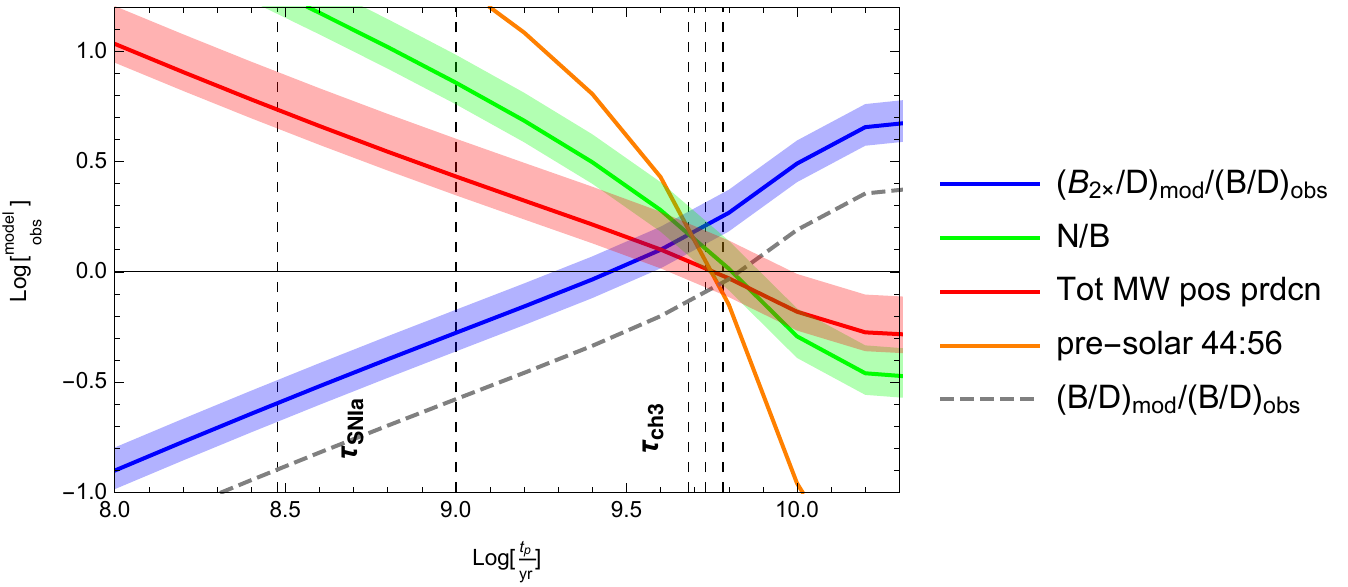}
\caption{{\bf Multiple constraints on the characteristic delay time $t_p$ of the main source of Galactic positrons.}
The curves and associated bands are identical to those shown in  main text Fig.~\ref{fig_plotCombinedConstraints} except the blue band labelled $B_{2 x}/D$ now 
shows the modelled, present-day ratio of bulge to disc positron sources divided by the observed ratio of bulge to disc positron luminosities ($0.42 \pm 0.09$[\cite{Siegert2015}])
as a function of $t_p$ under the assumption that, integrated over long timescales subsequent to star formation, 
the bulge is twice as efficient, per unit mass of stars formed, at forming positron sources (as might result, for instance, because of a top-heavy IMF in the bulge or an increased binary fraction
in the bulge).  
Such a scenario would suggest a very short $t_p \sim $2.5 Gyr, inconsistent with all the other curves.
This retrospectively justifies our assumption that there is a universal positron source
that is formed equally efficiently per unit mass of stars formed across the Galaxy. 
}
\label{fig_plotCombinedConstraintsSupplementaryInformation}
\end{figure}

In the main text we assumed that star formation is equally efficient (per unit mass of stars formed and integrating over long timescales post star formation) at generating positron sources.
Supplementary Fig.~\ref{fig_plotCombinedConstraintsSupplementaryInformation} demonstrates  that this conservative but strong assumption is supported by the evidence: if the bulge were twice as efficient at creating positron sources as the disc, a short delay time, inconsistent with the totality of other evidence, would be required.
This means that the differences in the positron injection rates between bulge, disc, and nucleus can be explained purely as a result of the different star formation histories of these regions (and the consequent differences in total stellar mass).
This finding is consistent with the facts that the metallicity distribution function of the disc and bulge are nearly the same\cite{Hayden2015,Ness2013} and 
that there is no convincing evidence for variations in initial mass function or binarity properties of these different environments at this time.

\subsection{Potential relevance of SN91bg scenario for the origin of Galactic positrons to SiC pre-solar `X-grains'}

It was noted some time ago\cite{Clayton1997} that 
the isotopically-anomalous, pre-solar SiC `X-grains' extracted from meteoritic material possess 
chemical and isotopic compositions resembling those naturally obtained from explosive helium burning. 
Moreover, such an environment provides a C/O
abundance ratio much greater than unity allowing carbon dust to grow without danger of oxidation;
in contrast, CC scenarios for the origin of X-grains suffer the generic problem that they require mixing of material from
different ejecta layers (where Si and other characteristic X-grain elements, including $^{44}$Ti, would be 
generated) separated by intervening levels that are O rich.
The scenario from ref.~\cite{Clayton1997}  envisaged He burning in a thin, $\sim 0.01 \msun$ cap atop a WD close to the Chandrasekhar limiting mass,
 i.e., likely an ordinary `single-degenerate' SN~Ia progenitor.
However, the subsequent non-observation (e.g., ref.~\cite{Gerardy2007}) of irrefutable dust condensation around ordinary SNe~Ia has called this model into question.

SN91bg are, in comparison to ordinary SNe~Ia, redder, cooler and more slowly expanding. 
Their ejecta, moreover, contain relatively less $^{56}$Ni, whose decay positrons act as a strong local heating source for SNe~Ia.
All these circumstances mean  that they are interesting candidates to host the production of dust and, in particular, SiC X-grains.
For efficient condensation, a total $^{28}$Si yield of $\sim 5 \times 10^{-4}$ per SN91bg[\cite{Waldman2011}], our current Galactic SN91bg rate yields a SiC production rate of a few $\times 10^{-5} \msun$/century, slightly more than the original model of ref.~\cite{ Clayton1997} and capable of sustaining the total Galactic SiC X-grain mass in steady state (modulo large uncertainties in dust destruction timescales).
Whether in fact the environment of a SN91bg does allow for the efficient condensation of X-grains is a subject for future theoretical and observational study.

\subsection{Arguments against rare events as sources of the Galactic $^{44}${Ti}}

Ref.~\cite{ The2006} raises two possible objections to the idea that rare, sub-luminous SN Ia events are the dominant source of $^{44}${Ti}:
\begin{enumerate}
\item
The lack of significant scatter in $^{44}${Ca}/$^{40}${Ca} in mainstream SiC grain abundances suggests
$^{44}$Ti-producing events are not rare.

\item The  $^{44}${Ca}/$^{48}${Ti} ratio in the so-called X-grains is never observed to be $\gsim 5$ whereas previous models of He burning\cite{ Clayton1997} suggests production in a ratio $\sim 100$.
\end{enumerate}
In fact, neither of these is insurmountable in the light of recent evidence:
Against 1), note that  ref.~\cite{ The2006} assumes that $^{44}$Ti is synthesised in ``$\alpha$-rich freeze-out of nuclear statistical equilibrium and secondarily from silicon burning" 
but this need not be true; in the scenario investigated by us it is actually synthesised in incomplete He burning, in which case roughly equal amounts of  
$^{44}${Ca} and $^{40}${Ca} are naturally produced.
Against 2) note that the nucleosynthesis yields obtained by ref.~\cite{ Clayton1997} referred to by ref.~\cite{ The2006} assume pure He composition.
More recent work\cite{ Waldman2011} shows, however, that allowing for small and astrophysically-plausible pollution by metals, can change yields of $^{48}$Ti very substantially.
In fact for model CO.5HE.2N.02 (a CO core of 0.5 $\msun$ and He envelope of 0.2 $\msun$ for which the envelope has a $^{14}$N  mass fraction 0.02 
corresponding to more-or-less solar metallicity) of ref.~\cite{ Waldman2011},
whereas the direct nucelsynthetic yield of $^{44}$Ti$/^{48}$Ti is $\sim 10^7$,
the yield of $^{48}$Cr (which decays via $^{48}$Cr $\to ^{48}$V $\to ^{48}$Ti in $<$20 days)
is $2 \times 10^{-2} \msun$ while the yield of $^{44}$Ti is $3 \times 10^{-2} \msun$.
Thus, the eventual, post-decay nuclear ratio is $^{44}${Ca}/$^{48}${Ti} $\simeq 0.7$ which easily matches
the constraint from the X-grains.
Also notable is that, for this model (which satisfies the $^{44}${Ca}/$^{48}${Ti} constraint for a realistic level of metal pollution in the He envelope) the absolute $^{44}$Ti yield is large and close to the values favoured by our analysis.

\subsection{Positrons from 2005E-like supernovae?}

One type of rare, sub-luminous thermonuclear supernova that probably involves He detonations capable of synthesising  $^{44}$Ti in the range suggested by Eq.~\ref{eqn_Min44TiMass} is
SN2005E-like.
Indeed refs.~\cite{ Perets2010,Perets2014} have already claimed SN2005E-like supernovae as potential sources of the Galactic positrons.
%
%Detailed calculations show, however, these events are too rare, given how much $^{44}$Ti they synthesise, and have the wrong delay time to supply the bulge positrons.
%
However, the
SN2005E-like population has a large  mean projected offset from host galaxies of 16 kpc[\cite{Yuan2013}] and seems to be associated with metal-poor stellar populations, implying they have the wrong distribution and are associated with the wrong sort of stars to account for a significant fraction of either the disc or bulge positrons.
A further problem is that they are too rare (for reasonable $^{44}$Ti yields) to supply the Galactic positrons\cite{Mernier2016}.

\subsection{Millisecond pulsars from accretion induced collapse and the GC Excess}

An interesting consequence of our picture 
is that it predicts a population of millisecond pulsars (MSPs) in the Galactic bulge (and, indeed, the disc)
whose progenitors are higher mass systems from the same binary WD population supplying (from its lower mass members) the SN91bg events. 
The progenitors of the MSPs are ONe-rich WDs  brought close to the Chandrasekhar mass by accretion from
(mostly) helium-burning star companions.
These undergo accretion induced collapse\cite{SI} into rapidly-spinning, magnetized neutron stars\cite{Ferrario2007,Hurley2010}.
This MSP population is plausibly responsible for the anomalous `GC Excess' (GCE)
$\gamma$-ray emission\cite{Hooper2011} that is coextensive\cite{Boehm2014} with the stellar bulge and the bulge positron signal.

Our binary population synthesis predicts 440,000 $(M_\mathrm{bulge})/(1.5 \times 10^{10} \ \msun)$ 
accretion induced collapse (AIC) events in the bulge over the life of the Galaxy.
In these events\cite{Jones2016} an ONeMg WD, accreting from a binary companion, is brought 
close to the Chandrasekhar mass increasing its central density  until electron captures on $^{20}$Ne and $^{24}$Mg occur. 
This releases sufficient energy to ignite $^{16}$O burning, which produces further isotopes favourable for electron capture
resulting, in particular regions of the parameter space, in catastrophic loss of electron degeneracy pressure and consequent gravitational collapse.
This results in a rapidly-spinning, magnetized neutron star;
such AIC events thus provide a route to the formation of MSPs
\cite{Michel1987,Ferrario2007,Hurley2010}.
For a death line (below which $\gamma$-ray emission ceases) at a $\gamma$-ray luminosity of $10^{33.5}$ erg s$^{-1}$[\cite{O'Leary2016}] we predict  from this, a population of $\sim 10^4$ extant MSPs
whose total bolometric emission is $\sim 4 \times 10^{37}$ erg s$^{-1}$, broadly matching GCE phenomenology (e.g., Ref.~\cite{Hooper2013}).

\subsection{Consistent $^{56}$Ni yields from different calculations}

\begin{figure}%[hb!]
\renewcommand{\thefigure}{S5}
\centering
\includegraphics[width = 1 \textwidth]{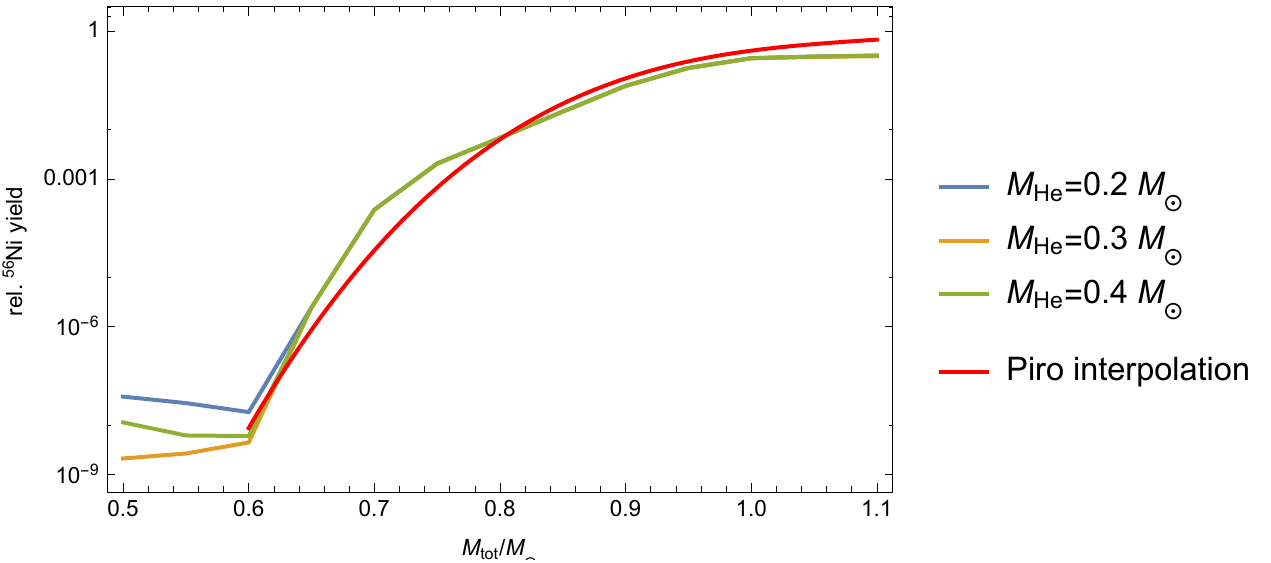}
\caption{{\bf Consistent} $^{56}$Ni {\bf yields from different calculations.} This figure demonstrates that the two calculations described in the Method section produce self-consistent results for the $^{56}$Ni yields of the He-CO WD merger product. }
\label{S5}
\end{figure}

Supplementary Fig.~\ref{S5} demonstrates that two different methods for determining the $^{56}$Ni yield of the He-CO WD merger product give consistent results that, for relevant He shell masses, scale purely with the total mass of the merger product.
As set out in the Methods section, in the first method, 
we produce an interpolation of the fractional $^{56}$Ni yield of CO burning as a function initial density as presented in fig.~A1 of ref.\cite{Fink2010}.
We then apply this parameterized yield to CO+He merger remnant density profiles generated by assuming that the remnants reach hydrostatic equilibrium before detonating.
For the second method, we use an existing interpolation\cite{Piro2014} of the modelled\cite{Sim2010} yield of $^{56}$Ni in detonations of single, sub-Chandrasekhar WDs that is given as a function purely of the total mass of the WD.

\subsection{Potential relevance of 1991bg-like supernovae to the galaxy cluster calcium anomaly}

Mernier et al.\cite{Mernier2016} recently examined
the X-ray derived elemental abundances in the hot intra-cluster medium (ICM) of 44 galaxy clusters, groups, and ellipticals.
They showed that numerical models of CC and ordinary SNe~Ia
yield abundances of $\alpha$-capture elements Ar, Ca, Cr, that 
tend to be systematically low when compared to the ICM abundances.
While it should be admitted that this deficit has a rather low statistical confidence, it may nevertheless be significant that, were it real, it could be rectified by allowing for a nucleosynthesis contribution by He detonations at long delay times.
Indeed, an overall best fit to the ICM abundance pattern is found\cite{Mernier2016} by incorporating putative He detonations yields
that are in line with 
model C0.5HE.2N.02 of ref.~\cite{Waldman2011}.
This model is for the detonation of a $0.2 \msun$ He shell with 0.02 mass fraction of $^{14}$N, overlaying a 0.5 $\msun$ CO core.
Taking the lead from Mulchaey et al.\cite{Mulchaey2014},
Mernier et al. identify these putative helium detonations with 
supernovae known as `Ca-rich gap transients' \cite{Filippenko2003,Perets2010}.
However, for the C0.5HE.2N.02 model, the best fit cluster abundance pattern is obtained for a rate of He detonations 34\% of the ordinary SN~Ia rate, much higher than the observed rate of Ca-rich gap transients.
For this reason, Mernier et al. dismiss the C0.5HE.2N.02 model despite the fact that it presents the best fit to the abundance data.
However: 
i) the best-fit rate of the He detonations of type C0.5HE.2N.02 relative to SNe~Ia is statistically identical to the rate of SN91bg relative to SNe~Ia in elliptical galaxies; 
ii) the yield of $^{44}$Ti in the C0.5HE.2N.02 model\cite{Waldman2011} is 0.03 $\msun$, consistent with our constraints on the Galactic $e^+$ source; and 
iii) in contrast to the ICM case, in pre-solar {\it elemental} data, the total Ca abundance (dominated by $^{40}$Ca) can be well explained without invoking an additional elemental Ca source like helium detonations.
Thus SNe91bg, synthesising $\sim 0.03 \msun$ of $^{44}$Ti, occurring at a rate today consistent with what we find above, and evolving according to a
long delay time (so as not to introduce too much total Ca into pre-solar system material), can  explain not only the Galactic positron population and the pre-solar $^{44}$Ca abundance, 
but also seem capable of addressing the abundance of elemental Ca (and likely Ar and Cr) in the ICM.

\subsection{Amplification/Recapitulation of argument that most Galactic positrons derive from a single type of stellar source}

We here recapitulate and amplify the arguments behind our contention that the totality of empirical data implies that there is a stellar  source, deriving from stars of age $\sim 5$ Gyr, that is the dominant source of antimatter in the Galaxy.

\begin{enumerate}
\item There are three distinct annihilation regions in the Galaxy: bulge, disc, and nucleus\cite{Siegert2015,Skinner2014}.

\item $\gamma$-ray constraints\cite{Aharonian1981,Beacom2006} on the injection kinetic energy of Galactic positrons imply that:	

\begin{enumerate}

\item $e^+$ cannot diffuse too far from their sources\cite{Alexis2014} so the annihilation radiation traces the positron source distribution (on $\gsim$ kpc scales) and the three distinct positron annihilation regions, therefore, should be identified with three distinct Galactic source distributions.
These three positron source regions of the Galaxy correspond to distinct stellar structures that have different star formation histories: whereas the nucleus and disc continue to form stars today (at $\sim 0.1$ and $\sim 1 \msun$ per year, respectively), the Galactic bulge has experienced negligible star formation since 8-10 Gyr ago\cite{Nataf2015}.

\item Many dark matter and/or compact object and/or diffuse cosmic ray origin scenarios for the positrons are ruled out or, at least, disfavoured by the injection energy 	constraints.
Moreover, a recent paper\cite{Wilkinson2016} claims to exclude  light, thermally-produced WIMPs as an explanation of the bulge positrons on the basis of cosmological data. 
(This would rule out one of the previous strongest contenders for a dark matter explanation of the bulge positrons.)

\end{enumerate}

\item To explain the global positron annihilation phenomenology of the Galaxy, the broad possibilities for the positron sources are that there are either

\begin{enumerate}

\item Stellar source(s) in the disc and `special' (i.e., non stellar) positron source(s) in the bulge/nucleus. Note that 
there are old stars in both disc and bulge (these old star populations having similar total mass) but no recent star formation in the bulge.
Therefore
the difference between the bulge and disc stellar populations hangs on the possession of stars age $\lsim8$ Gyr. 
Thus this general scenario seems to require that stars of age $\lsim8$ Gyr in the disc make disc positrons and something else makes bulge positrons; we describe a number of problems with such a scenario immediately below.

\item  Alternatively,
the same type/distribution of source(s) everywhere.

\end{enumerate}

\end{enumerate}

\noindent
Some top-level objections to a `special source' scenario as described in 3(a) above:

\begin{enumerate}[I.]

\item It becomes then a striking, unexplained coincidence that the positron luminosity ratio of both the bulge/disc and the nucleus/bulge is equal to the stellar mass ratios of these structures.

\item More generally, a fine tuning is implied: why should the bulge and disc positron signals be so similar in amplitude if they derive from completely unrelated processes?

%\item It is also then unexplain why the annihilation phenomenology (511 keV line width, positronium fraction, etc) is observationally very similar\cite{Siegert2015} between the bulge and disc environments if the 

\item Another problem: the difficulty of finding an acceptable `special source' in the bulge. Candidates for this source: 

\begin{enumerate}

\item A bulge positron source located in the nucleus?  Diffusion from the nucleus out to the full $\sim$ 3 kpc size of the bulge seems to be excluded\cite{Alexis2014} so the positron transport would have to be in a nuclear wind. It is not yet established whether the observed limit to the Doppler shift of the bulge annihilation line signal is consistent with the required wind velocity. Perhaps the super-massive black hole could supply the positrons but the question of whether it can deliver positrons which annihilate in the observed ratio between nucleus and bulge has not been addressed. Non-trivially, such positrons would also have to obey the injection energy constraints. Alternatively, nuclear star formation in the Central Molecular Zone perhaps injects positrons that are subsequently advected to the bulge.
In this case, note that the nuclear star formation rate is $\sim 10$\% of the disc rate.
We would therefore need to explain how nuclear star formation can be approximately an order of magnitude more efficient in producing positrons with respect to the disc star formation.

\item {\it In situ} injection of bulge positrons by dark matter? This seems very unlikely, as explained already, if not formally excluded. 
\end{enumerate}

\item Yet another problem: difficulty finding an acceptable stellar positron source (and $\beta^+$-unstable radionuclide) in the disc associated with stars of age $\lsim 8$ Gyr:

\begin{enumerate}

\item Ordinary SNe~Ia seem to satisfy this age constraint on the stellar progenitors but, as we explain in the main article, positrons from $^{56}$Ni do not seem to escape SN~Ia ejecta. 

\item $^{26}$Al only supplies $\sim$10\% of the total Galactic positron luminosity given the intensity of the 1.8 MeV $^{26}$Al line (whose intensity is in steady state). Moreover the disc annihilation emission seems to require a geometric model significantly more extended in latitude and longitude than areas of recent star formation/massive stars in disc (which, in fact, seem to be well traced by the 1.8 MeV $\gamma$-ray line from $^{26}$Al\cite{Prantzos2011}). Positron transport might plausibly explain the $\sim$ kpc latitudinal extent of the disc annihilation geometric model (which, note, actually matches the thick disc of old stars similar in age to the bulge) but not the longitudinal extent. In short, the disc annihilation geometry directly suggests an old stellar population\cite{Bouchet2010}, likely in tension with the $\lsim 8$ Gyr age requirement of the special source scenario.

\item $^{44}$Ti positron sources have to be infrequent (to evade the constraint implied by the dearth of $^{44}$Ti gamma-ray/X-ray line emission in the Galaxy) and (consequently) need to yield large amounts of $^{44}$Ti per event as already outlined. Such large $^{44}$Ti yields require explosive He burning at a high density of $\sim 10^{5-6}$ g cm$^{-3}$; the assembly of large amounts of reasonably pure He in conditions suitable for detonation points to low mass stars that do not experience core He burning -- but then that implies long main sequence lifetimes, which is again in potential tension with the $\lsim 8$ Gyr age requirement.

\end{enumerate}

\end{enumerate}

In summary of the above, there are many points telling against the scenario of a `special' positron source in the bulge and a different, stellar positron source in disc.
In contrast, the alternative explanation that we have investigated in this paper -- that there is a single, dominant, class of stellar positron sources in the Galaxy -- evades all these arguments. This class is necessarily related to old stars because the stellar population of the bulge is exclusively old.

%%%%%%%%%%%%%%%%%%%%%%%%%%%%%%%%%%%%%%%%%%%%%%%%%%%%%%%%%%%%%%%%%%%%%%%%%%%%%%%

\renewcommand\refname{Supplementary Information References}
{}

%%%%%%%%%%%%%%%%%%%%%%%%%%%%%%%%%%%%%%%%%%%%%%%%%%%%%%%%%%%%%%%%%%%%%%%%%%%%%%%


\begin{thebibliography}{}

\bibitem%[Johnson et al.(1972)]
{Johnson1972} Johnson, W.~N., III, Harnden, F.~R., Jr., \& Haymes, R.~C. The Spectrum of Low-Energy Gamma Radiation from the Galactic-Center Region \apjl, {\bf 172}, 1-7 (1972)

\bibitem%[Prantzos et al.(2011)]
{Prantzos2011} Prantzos, N., Boehm, 
C., Bykov, A.~M., et al. The 511 keV emission from positron annihilation in the Galaxy {\it Rev.~Mod.~Phys.}, {\bf 83}, 1001-1056 (2011)

\bibitem{SI}
See the Supplementary Information.

\bibitem%[Bland-Hawthorn \& Gerhard(2016)]
{Bland-Hawthorn2016} Bland-Hawthorn, J., \& Gerhard, O. The Galaxy in Context:
Structural, Kinematic,
and Integrated Properties \araa, {\bf 54}, 529-596  (2016)

\bibitem%[Aharonian \& Atoyan(1981)]
{Aharonian1981} Aharonian, F.~A., \& Atoyan, A.~M. 	
On the origin of the galactic annihilation radiation {\it Sov.~Astr.~Let.}, {\bf 7}, 395-398  (1981)

\bibitem%[Beacom \& Y{\" u}ksel(2006)]
{Beacom2006} Beacom, J.~F., Y{\" u}ksel, H. Stringent Constraint on Galactic Positron Production \prl, {\bf 97}, 071102, 1-4  (2006)

\bibitem%[Alexis et al.(2014)]
{Alexis2014} Alexis, A., Jean, P., Martin, P., \& Ferri{\`e}re, K. Monte Carlo modelling of the propagation and annihilation of nucleosynthesis positrons in the Galaxy \aap, {\bf 564}, A108  1-14 (2014)

\bibitem%[Churazov et al.(2011)]
{Churazov2011} Churazov, E., Sazonov, 
S., Tsygankov, S., Sunyaev, R., 
\& Varshalovich, D. Positron annihilation spectrum from the Galactic Centre region observed by SPI/INTEGRAL revisited: annihilation in a cooling ISM? \mnras, {\bf 411}, 1727-1743  (2011)

\bibitem%[Siegert et al.(2016)]
{Siegert2015} Siegert, T., Diehl, R., Khachatryan, G., et al. Gamma-ray spectroscopy of positron annihilation in the Milky Way
 \aap, {\bf 586}, A84 1-16 (2016)

\bibitem%[Roques et al.(2003)]
{Roques2003} Roques, J.~P., Schanne, S., von Kienlin, A., et al. SPI/INTEGRAL in-flight performance \aap, {\bf 411}, L91-100  (2003)

\bibitem%[Skinner et al.(2014)]
{Skinner2014} 
Skinner, G., Diehl, R., Zhang, X., Bouchet, L., \& Jean, P., in Proceedings
of the 10th INTEGRAL Workshop: ÓA Synergistic View of the High-Energy SkyÓ (INTEGRAL 2014). 15-19 September 2014. Annapolis, MD, USA. Published online at http://pos.sissa.it/cgi-bin/reader/conf.cgi?confid=228, id.054, {\bf 054}  (2014)

\bibitem%[Launhardt et al.(2002)]
{Launhardt2002} Launhardt, R., Zylka, R., \& Mezger, P.~G. The nuclear bulge of the Galaxy
Astronomy III. Large-scale physical characteristics of stars and interstellar matter \aap, {\bf 384}, 112-138  (2002)


\bibitem%[Higdon et al.(2009)]
{Higdon2009} Higdon, J.~C., 
Lingenfelter, R.~E., \& Rothschild, R.~E. The Galactic Positron Annihilation Radiation \& The Propagation of Positrons in the Interstellar Medium \apj, {\bf 698}, 350-379 (2009)

\bibitem%[Guessoum et al.(2006)]
{Guessoum2006} Guessoum, N., Jean, P., \& Prantzos, N. Microquasars as sources of positron annihilation radiation \aap, {\bf 457}, 753-762  (2006)

\bibitem%[Siegert et al.(2016)]
{Siegert2016} Siegert, T., Diehl, R., Greiner, J., et al. Positron annihilation signatures associated with the outburst of the microquasar V404 Cygni \nat, {\bf 531}, 341-343  (2016)

\bibitem%[Chan \& Lingenfelter(1993)]
{Chan1993} Chan, K.-W., \& Lingenfelter, R.~E. Positrons from supernovae \apj, {\bf 405}, 614-636  (1993)

\bibitem%[Leloudas et al.(2009)]
{Leloudas2009} Leloudas, G., Stritzinger, M.~D., Sollerman, J., et al. The normal Type Ia SN 2003hv out to very late phases \aap, {\bf 505}, 265-279  (2009)

\bibitem%[Kerzendorf et al.(2014)]
{Kerzendorf2014} Kerzendorf, W.~E., 
Taubenberger, S., Seitenzahl, I.~R., 
\& Ruiter, A.~J. Very Late Photometry of SN 2011fe \apjl, {\bf 796}, 26-30  (2014)

\bibitem%[Graur et al.(2016)]
{Graur2016} Graur, O., Zurek, D., 
Shara, M.~M., et al. Late-time Photometry of Type Ia Supernova SN 2012cg Reveals the Radioactive Decay of 57 Co \apj, {\bf 819}, 31, 1-8  (2016)

\bibitem%[Bouchet et al.(2015)]
{Bouchet2015} Bouchet, L., Jourdain, 
E., \& Roques, J.-P. The Galactic $^{26}$Al Emission Map as Revealed by INTEGRAL SPI \apj, {\bf 801}, 142, 1-15  (2015)

\bibitem%[Troja et al.(2014)]
{Troja2014} Troja, E., Segreto, A., La Parola, V., et al. Swift/BAT Detection of Hard X-Rays from Tycho's Supernova Remnant: Evidence for Titanium-44 \apjl, {\bf 797}, 6-10  (2014)

\bibitem%[The et al.(2006)]
{The2006} The, L.-S., Clayton, D.~D., Diehl, R., et al. Are $^{44}$Ti-producing supernovae exceptional? \aap, {\bf 450}, 1037-1050  (2006)

\bibitem%[Timmes et al.(1996)]
{Timmes1996} Timmes, F.~X., Woosley, 
S.~E., Hartmann, D.~H., \& Hoffman, R.~D. The Production of $^{44}$Ti and $^{60}$Co in Supernovae \apj, {\bf 464}, 332-341  (1996)

\bibitem%[Leising \& Share(1994)]
{Leising1994} Leising, M.~D., \& Share, G.~H. Gamma-Ray limits on Galactic $^{60}$Fe and $^{44}$Ti nucleosynthesis, \apj, {\bf 424}, 200-207  (1994)

\bibitem%[Hansen(1971)]
{Hansen1971} Hansen, C.~J. Element Production in Simple Helium Burning \apj, {\bf 169}, 
585-588  (1971)

\bibitem%[Woosley et al.(1986)]
{Woosley1986} Woosley, S.~E., Taam, R.~E., \& Weaver, T.~A. Models for Type I supernova. I - Detonations in white dwarfs \apj, {\bf 301}, 601-623  (1986)

\bibitem%[Karakas et al.(2015)]
{Karakas2015} Karakas, A.~I., Ruiter, A.~J., \& Hampel, M. R Coronae Borealis Stars Are Viable Factories of Pre-solar Grains \apj, {\bf 809}, 184, 1-6  (2015)

\bibitem%[Belczynski et al.(2008)]
{Belczynski2008} Belczynski, K., Kalogera, V., Rasio, F.~A., et al. Compact Object Modeling with the StarTrack Population Synthesis Code \apjs, {\bf 174}, 223-260  (2008)

\bibitem%[Pakmor et al.(2013)]
{Pakmor2013} Pakmor, R., Kromer, M., 
Taubenberger, S., \& Springel, V. Helium-ignited Violent Mergers as a Unified Model for Normal and Rapidly Declining Type Ia Supernovae
 \apjl, {\bf 770}, 8, 1-7  (2013)

\bibitem%[Dan et al.(2015)]
{Dan2015} Dan, M., Guillochon, J., Br{\"u}ggen, M., Ramirez-Ruiz, E., \& Rosswog, S. Thermonuclear detonations ensuing white dwarf mergers
 \mnras, {\bf 454}, 4411Ð4428  (2015)

\bibitem%[Filippenko et al.(1992)]
{Filippenko1992} Filippenko, A.~V., Richmond, M.~W., Branch, D., et al. The subluminous, spectroscopically peculiar type IA supernova 1991bg in the elliptical galaxy NGC 4374 \aj, {\bf 104}, 1543-1556  (1992)

\bibitem%[Perets et al.(2010)]
{Perets2010} Perets, H.~B., Gal-Yam, 
A., Mazzali, P.~A., et al. A faint type of supernova from a white dwarf with a helium-rich companion \nat, {\bf 465}, 322-325  (2010)

\bibitem%[Li et al.(2011a)]
{Li2011a} Li, W., Leaman, J., Chornock, R., 
et al. Nearby supernova rates from the Lick Observatory Supernova Search - III. The rate-size relation, and the rates as a function of galaxy Hubble type and colour \mnras, {\bf 412}, 1473-1507  (2011)
 
\bibitem%[Howell(2001)]
{Howell2001} Howell, D.~A. The Progenitors of Subluminous Type Ia Supernovae \apjl, 
{\bf 554}, 193-196  (2001)

\bibitem%[Gonz{\'a}lez-Gait{\'a}n et al.(2011)]
{Gonzalez-Gaitan2011} 
Gonz{\'a}lez-Gait{\'a}n, S., Perrett, K., Sullivan, M., et al. Subluminous Type Ia Supernovae at High Redshift from the Supernova Legacy Survey \apj, {\bf 727}, 107, 1-18  (2011)


\bibitem%[Piro et al.(2014)]
{Piro2014} Piro, A.~L., Thompson, 
T.~A., \& Kochanek, C.~S. Reconciling $^{56}$Ni production in Type Ia supernovae with double degenerate scenarios \mnras, {\bf 438}, 3456-3464  (2014)


\bibitem%[van Kerkwijk et al.(2010)]
{vanKerkwijk2010} van Kerkwijk, M.~H., Chang, P., \& Justham, S. Sub-Chandrasekhar White Dwarf Mergers as the Progenitors of Type Ia Supernovae \apjl, {\bf 722}, 157-161  (2010)


\bibitem%[Sullivan et al.(2011)]
{Sullivan2011} Sullivan, M., Kasliwal, M.~M., Nugent, P.~E., et al. The Subluminous and Peculiar Type Ia Supernova PTF 09dav \apj, {\bf 732}, 118, 1-13  (2011)

\bibitem%[Nataf(2015)]
{Nataf2015} Nataf, D.~M. The Controversial Star-Formation History and Helium Enrichment of the Milky Way Bulge,
{\it P.~Astron.~Soc.~Aust.},  {\bf 33}, e023 1-7 (2016)

\bibitem%[Childress et al.(2014)]
{Childress2014} Childress, M.~J., Wolf, C., \& Zahid, H.~J. Ages of Type Ia supernovae over cosmic time \mnras, {\bf 445}, 1898-1911 (2014)

\bibitem%[Holcomb et al.(2013)]
{Holcomb2013} Holcomb, C., 
Guillochon, J., De Colle, F., \& Ramirez-Ruiz, E. Conditions for Successful Helium Detonations in Astrophysical Environments \apj, {\bf 771}, 14, 1-8  (2013)

\bibitem%[Sim et al.(2010)]
{Sim2010} Sim, S.~A., R{\"o}pke, F.~K., Hillebrandt, W., et al. Detonations in Sub-Chandrasekhar-mass C+O White Dwarfs \apjl, {\bf 714}, 52-57  (2010)

\bibitem%[Snaith et al.(2014)]
{Snaith2014} Snaith, O.~N., Haywood, M., Di Matteo, P., et al. The Dominant Epoch of Star Formation in the Milky Way Formed the Thick Disk \apjl, {\bf 781}, 31, 1-5  (2014)

\bibitem%[Lodders(2003)]
{Lodders2003} Lodders, K. Solar System Abundances and Condensation Temperatures of the Elements \apj, {\bf 591}, 1220-1247  (2003)


\bibitem%[Waldman et al.(2011)]
{Waldman2011} Waldman, R., Sauer, D., 
Livne, E., et al. Helium Shell Detonations on Low-mass White Dwarfs as a Possible Explanation for SN 2005E \apj, {\bf 738}, 21, 1-12  (2011)



\end{thebibliography}

\begin{thebibliography}{}
\setcounter{enumiv}{45}

\bibitem%[van Dokkum et al.(2013)]
{vanDokkum2013} van Dokkum, P.~G., Leja, J., Nelson, E.~J., et al. The Assembly of Milky-Way-like Galaxies Since z $\sim$ 2.5 \apjl, {\bf 771}, 35, 1-7 (2013)

\bibitem%[Licquia \& Newman(2015)]
{Licquia2015} Licquia, T.~C., \& Newman, J.~A., Improved Estimates of the Milky Way's Stellar Mass and Star Formation Rate from Hierarchical Bayesian Meta-Analysis \apj, {\bf 806}, 96, 1-20  (2015)

\bibitem%[Krumholz.~et al.(2016)]
{Krumholz2016} Krumholz., M.~R., Kruijssen, J.~M.~D., \& Crocker, R.~M. A dynamical model for gas flows, star formation and nuclear winds in galactic centres. \mnras {\bf 466},  1213-1233  (2017)

\bibitem%[Figer et al.(2004)]
{Figer2004} Figer, D.~F., Rich, R.~M., Kim, S.~S., Morris, M., \& Serabyn, E. An Extended Star Formation History for the Galactic Center from Hubble Space Telescope NICMOS Observations \apj, {\bf 601}, 319-339  (2004)

\bibitem%[Ruiter et al.(2009)]
{Ruiter2009} Ruiter, A.~J., Belczynski, K., \& Fryer, C. Rates and Delay Times of Type Ia Supernovae \apj, {\bf 699}, 2026-2036  (2009)


\bibitem%[Foreman-Mackey et al.(2013)]
{Foreman-Mackey2013} Foreman-Mackey, D., Hogg, D.~W., Lang, D., \& Goodman, J. 	
emcee: The MCMC Hammer \pasp, {\bf 125}, 306  (2013)

\bibitem%[Fink et al.(2010)]
{Fink2010} Fink, M., R{\"o}pke, F.~K., Hillebrandt, W., et al. Double-detonation sub-Chandrasekhar supernovae: can minimum helium shell masses detonate the core? \aap, {\bf 514}, A53, 1-10  (2010)

\bibitem%[Moore et al.(2013)]
{Moore2013} Moore, K., Townsley, D.~M., \& Bildsten, L. The Effects of Curvature and Expansion on Helium Detonations on White Dwarf Surfaces \apj, {\bf 776}, 97, 1-21  (2013)


\end{thebibliography}

\begin{thebibliography}{}
%\renewcommand{\refname}{Supplementary Information References}
\setcounter{enumiv}{52}


\bibitem%[Bouchet et al.(2010)]
{Bouchet2010} Bouchet, L., Roques, J.~P., \& Jourdain, E.\ 2010, \apj, 720, 1772 

\bibitem%[Minniti et al.(2010)]
{Minniti2010} Minniti, D., Lucas, P.~W., Emerson, J.~P., et al.\ 2010, New Astronomy, 15, 433 


%\bibitem%[Howell et al.(2001)]
%{Howell2001b} Howell, D.~A., 
%H{\"o}flich, P., Wang, L., \& Wheeler, J.~C.\ 2001, \apj, 556, 302 

%\bibitem%[Patat et al.(2012)]
%{Patat2012} Patat, F., H{\"o}flich, P., Baade, D., et al.\ 2012, \aap, 545, A7 

%\bibitem%[Pakmor et al.(2011)]
%{Pakmor2011} Pakmor, R., Hachinger, S., R{\"o}pke, F.~K., \& Hillebrandt, W.\ 2011, \aap, 528, A117 

\bibitem%[Ruiter et al.(2013)]
{Ruiter2013} Ruiter, A.~J., Sim, S.~A., Pakmor, R., et al.\ 2013, \mnras, 429, 1425 


\bibitem%[Yuan et al.(2013)]
{Yuan2013} Yuan, F., Kobayashi, C., 
Schmidt, B.~P., et al.\ 2013, \mnras, 432, 1680 

\bibitem%[Jones et al.(2016)]
{Jones2016} Jones, S., Roepke, F.~K., Pakmor, R., et al.\ 2016, arXiv:1602.05771 

\bibitem%[O'Leary et al.(2016)]
{O'Leary2016} O'Leary, R.~M., Kistler, M.~D., Kerr, M., \& Dexter, J.\ 2016, arXiv:1601.05797 

\bibitem%[Hooper \& Slatyer(2013)]
{Hooper2013} Hooper, D., \& Slatyer, T.~R.\ 2013, Physics of the Dark Universe, 2, 118 

\bibitem%[Perets(2014)]
{Perets2014} Perets, H.~B.\ 2014, 
arXiv:1407.2254 

\bibitem%[Ness et al.(2013)]
{Ness2013} Ness, M., Freeman, K., Athanassoula, E., et al.\ 2013, \mnras, 430, 836 

\bibitem%[Hayden et al.(2015)]
{Hayden2015} Hayden, M.~R., Bovy, J., Holtzman, J.~A., et al.\ 2015, \apj, 808, 132 

\bibitem%[Michel(1987)]
{Michel1987} Michel, F.~C.\ 1987, \nat, 329, 310 

\bibitem%[Ferrario \& Wickramasinghe(2007)]
{Ferrario2007} Ferrario, L., \& Wickramasinghe, D.\ 2007, \mnras, 375, 1009 

\bibitem%[Hurley et al.(2010)]
{Hurley2010} Hurley, J.~R., Tout, C.~A., Wickramasinghe, D.~T., Ferrario, L., \& Kiel, P.~D.\ 2010, \mnras, 402, 1437 

\bibitem%[Hooper \& Goodenough(2011)]
{Hooper2011} Hooper, D., \& Goodenough, L.\ 2011, Physics Letters B, 697, 412 

\bibitem%[Boehm et al.(2014)]
{Boehm2014} Boehm, C., Gondolo, P., Jean, P., et al.\ 2014, arXiv:1406.4683 

\bibitem%[Clayton et al.(1997)]
{Clayton1997} Clayton, D.~D., Arnett, 
D., Kane, J., \& Meyer, B.~S.\ 1997, \apj, 486, 824 

\bibitem%[Gerardy et al.(2007)]
{Gerardy2007} Gerardy, C.~L., Meikle, W.~P.~S., Kotak, R., et al.\ 2007, \apj, 661, 995 

\bibitem%[Guessoum et al.(2010)]
{Guessoum2010} Guessoum, N., Jean, P., \& Gillard, W.\ 2010, \mnras, 402, 1171 

\bibitem%[Martin et al.(2012)]
{Martin2012} Martin, P., Strong, A.~W., Jean, P., Alexis, A., \& Diehl, R.\ 2012, \aap, 543, A3 

\bibitem%[Dufour \& Kaspi(2013)]
{Dufour2013} Dufour, F., \& Kaspi, V.~M.\ 2013, \apj, 775, 52 

\bibitem%[Mernier et al.(2016)]
{Mernier2016} Mernier, F., de Plaa, J., Pinto, C., et al.\ 2016, \aap, 595, A126 

\bibitem%[Mulchaey et al.(2014)]
{Mulchaey2014} Mulchaey, J.~S., Kasliwal, M.~M., \& Kollmeier, J.~A.\ 2014, \apjl, 780, L34 

\bibitem%[Filippenko et al.(2003)]
{Filippenko2003} Filippenko, A.~V., Chornock, R., Swift, B., et al.\ 2003, IAU Circular 8159, 2 

\bibitem%[Wilkinson et al.(2016)]
{Wilkinson2016} Wilkinson, R.~J., Vincent, A.~C., Boehm, C., \& McCabe, C.\ 2016, arXiv:1602.01114 

\end{thebibliography}
\end{document}